\documentclass[useAMS,usenatbib,a4paper]{mn2e}
\usepackage{epsf,times}
\usepackage{amsmath}
\usepackage{amssymb}
\usepackage{epsfig}
\usepackage[draft]{hyperref}
\usepackage{wasysym}
\usepackage{tipa}
\usepackage{rotating}
\usepackage{color}
\usepackage[total={17.8cm,24.0cm},centering]{geometry} 
\newcommand{\etal}{{et al}\/.}

\begin{document}
\title[{\rm Herschel}-ATLAS: Radio-selected galaxies]{{\it Herschel}-ATLAS/GAMA: 
What determines the far-infrared properties of radio-galaxies?*}
\author[J.S.\ Virdee \etal]
{J.S.\ Virdee$^{1,2}$,
M.J.\ Hardcastle$^3$,
S.\ Rawlings$^1$,
D.\ Rigopoulou$^{1,2}$,
T. Mauch$^{3,4}$,
M.J.\ Jarvis$^{1,5}$,
\newauthor
A. Verma$^1$,
D.J.B.\ Smith$^3$,
I.\ Heywood$^1$,
S.V. White$^1$,
M. Baes$^6$,
A. Cooray$^{7,8}$,
G. De Zotti$^{9,10}$,
\newauthor
S. Eales$^{11}$, 
M. J. Micha{\l}owski$^{12}$,
N. Bourne$^{13}$,
A. Dariush$^{14}$,
L. Dunne$^{15}$, 
R. Hopwood$^{16,17}$,
\newauthor
E. Ibar$^{18,19}$,
S. Maddox$^{15}$,
M. W. L. Smith$^{11}$ \&
E. Valiante$^{11}$
\\
$^1$ Oxford Astrophysics, Denys Wilkinson Building, University of Oxford, 
Keble Rd, Oxford OX1 3RH\\
$^{2}$ Space Science \& Technology Department, Rutherford Appleton 
Laboratory, Chilton, Didcot, Oxfordshire OX11 0QX\\
$^3$ School of Physics, Astronomy and Mathematics, University of
Hertfordshire, College Lane, Hatfield AL10 9AB\\
$^4$SKA Africa, 3rd Floor, The Park, Park Road, Pinelands, 7405, South Africa\\
$^5$ Physics Department, University of the Western Cape, Cape Town, 
7535, South Africa\\
$^{6}$ Sterrenkundig Observatorium, Universiteit Gent, Krijgslaan 281 S9, B-9000 Gent, Belgium\\
$^{7}$ Department of Physics and Astronomy, University of California, Irvine, 
CA 92697, USA\\
$^{8}$ California Institute of Technology, 1200 E. California Blvd, Pasadena, 
CA 91125, USA\\
$^{9}$ INAF-Osservatorio Astronomico di Padova, Vicolo dell'Osservatorio 5, 
I-35122, Padova, Italy\\
$^{10}$ Astrophysics Sector, SISSA, Via Bonomea 265, I-34136 Trieste, Italy\\
$^{11}$ School of Physics \&\ Astronomy, Cardiff University, The Parade, Cardiff, CF24 3AA\\
$^{12}$ Scottish Universities Physics Alliance, Institute for Astronomy, University of Edinburgh, Royal Observatory, Edinburgh, EH9 3HJ\\
$^{13}$ School of Physics \&\ Astronomy, University of Nottingham, Nottingham NG7 2RD\\
$^{14}$ Institute of Astronomy, University of Cambridge, Madingley Road, Cambridge, CB3 0HA, United Kingdom\\
$^{15}$ Department of Physics and Astronomy, University of Canterbury, Private Bag 4800, 
Christchurch 8140, New Zealand.\\
$^{16}$ Department of Physics and Astronomy, The Open University,
Walton Hall, Milton Keynes, MK7 6AA\\
$^{17}$ Astrophysics Group, Blackett Lab, Imperial College London,
Prince Consort Road, London SW7 2AZ\\
$^{18}$ UK Astronomy Technology Centre, The Royal Observatory, Blackford Hill, Edinburgh, EH9 3HJ, UK\\
$^{19}$ Universidad Cat\'olica de Chile, Departamento de Astronom\'ia y Astrof\'isica, Vicu\~na Mackenna 4860, Casilla 306, Santiago 22, Chile\\
\\
\\
Submitted 23 October 2012\\
Accepted 18 March 2013\\
\\
\\
\\
\\
\\
\\
\\
\\
\\
\\
\\
\\
\\
\\
\\
\\
\\
\\
\\
\\
\\
\\
\\
\\
\\
\\
\\
\\
\\
\\
\\
\\
\\
}
\maketitle
\vbox{\vskip -60pt}
%
%
\begin{abstract}

We perform a stacking analysis of \emph{Herschel}-ATLAS data in order to 
obtain isothermal dust temperatures and rest-frame luminosities at 250\,$\mu$m 
($L_{250}$), for a well-defined sample of 1599 radio sources over the H-ATLAS 
Phase 1\,/\,Galaxy and Mass Assembly (GAMA) area. The radio sample is generated 
using a combination of NRAO VLA Sky Survey (NVSS) data and K-band UKIDSS - 
Large Area Survey (LAS) data, over the redshift range $0.01<z<0.8$. The FIR properties 
of the sample are investigated as a function of 1.4-GHz luminosity, redshift, 
projected radio-source size and radio spectral index. In order to search for stellar-mass
dependent relations, we split the parent sample into those sources which are 
below and above 1.5\,$L_{K}^{*}$.

After correcting for stellar mass and redshift, we find no relation between the 
250-$\mu$m luminosity and the 1.4-GHz radio luminosity of radio AGN. 
This implies that a galaxy's nominal radio luminosity has little or no bearing 
on the star-formation rate and\,/\,or dust mass content of the host system, although 
this does not mean that other variables (e.g. radio source size) related to the jets 
do not have an effect.
The $L_{250}$ of both the radio detected and non radio-detected galaxies (defined 
as those sources not detected at 1.4-GHz but detected in the Sloan Digital Sky Survey with 
$r'<22$) rises with increasing redshift. For matched samples in $L_{K}$ and g$'$-\,r$'$, 
sub-1.5\,$L_{K}^{*}$ and super-1.5\,$L_{K}^{*}$ radio-detected galaxies have 0.89$\pm$0.18 
and 0.49$\pm$0.12 times the 250\,$\mu$m luminosity of their non radio-detected 
counterparts. Thus, while no difference in $L_{250}$ is observed in sub-1.5\,$L_{K}^{*}$ 
radio-detected galaxies, a strong deficit is observed in super-1.5\,$L_{K}^{*}$ radio-detected 
galaxies. We explain these results in terms of the hotter, denser and richer halo
environments massive radio-galaxies maintain and are embedded in. These 
environments are expected to quench the cold gas and dust supply needed for 
further star-formation and therefore dust production. Our results indicate that
all massive radio galaxies ($>1.5$\,$L_{K}^{*}$) may have systematically lower 
FIR luminosities ($\sim25$ per cent) than their colour-matched non radio-detected 
counterparts.

Compact radio sources ($<$\,30\,kpc) are associated with higher 250\,$\mu$m 
luminosities and dust temperatures than their more extended ($>$\,30\,kpc) counterparts. 
The higher dust temperature suggests that this may be attributed to enhanced 
star-formation rates (SFR) in compact radio galaxies, but whether this is directly or 
indirectly due to radio activity (e.g. jet induced or merger-driven star formation) is as 
yet unknown. Finally, no relation between radio spectral index and $L_{250}$ is 
found for the subset of 1.4-GHz radio sources with detections at 330\,MHz. 

\end{abstract}


\begin{minipage}{17.5cm}
{\footnotesize * {\it Herschel} is an ESA space observatory with science instruments
provided by European-led Principal Investigator consortia and with
important participation from NASA.}
\end{minipage}


\clearpage
\section{Introduction}
\label{intro}

Models which seed and evolve galaxies in accordance with the growth of their 
dark matter haloes lead to an over-prediction of both the number of faint-end 
(low luminosity) and bright-end (high luminosity) optical galaxies in the local 
Universe (e.g. \citealt{Springel_2005}). 
The faint-end problem has largely been resolved by the introduction into the 
models of photo-ionization of the pre-galactic gas and supernova feedback 
(e.g. \citealt{Cole_2000}; \citealt{Benson_2003}). The 
net effect is to eject gas and dust from small dark matter halos, accounting 
for the relative paucity of low-luminosity galaxies. However, the 
bright-end problem remains; the ejected gas in systems embedded in 
higher-mass halos cannot escape and therefore cools back onto the 
central galaxy, stimulating star formation and by extension, galaxy growth. 
An additional source of energy is required either to eject the gas or prevent
its accretion onto the galaxy in the first place. This motivates the need to tap into 
the vast energy released by the accretion of gas onto central, supermassive 
black-holes (e.g. \citealt{Granato_2004}; \citealt{DiMatteo_2005}; \citealt{Bower_2006}; 
\citealt{Croton_2006}). 

There is good circumstantial evidence that active galactic nuclei (AGN) 
activity is related to galaxy growth. The increase in the star formation density of the Universe with 
redshift (e.g.. \citealt{Madau_1996}; \citealt{Hopkins_Beacom_2006}) is mirrored 
by an increase in the luminosity density of quasars (e.g. \citealt{Boyle_1998}; 
\citealt{Richards_2006}; \citealt{Croom_2009}), with both peaking at $z\sim2$; 
the correlation between galaxy-bulge mass and black-hole mass 
(e.g. \citealt{Magorrian_1998}; \citealt{Marconi_2003}; \citealt{Haring_2004}) 
is suggestive of a strong physical link. However, the nature of these relationships 
is much less clear: does AGN activity directly stimulate, regulate and finally terminate 
galaxy wide star formation and, if so, what types of AGN activity do this? Or can a 
common phenomenon trigger both star-formation and black-hole accretion 
leading to simultaneous but largely independent activity
(e.g. major mergers: \citealt{Granato_2004}; \citealt{Di_Matteo_2005})?
Direct observation of the star-forming properties of large numbers of AGN
will be crucial in order to answer such questions.

Broadly, AGN may be split into two groups defined by their radiative efficiency.
In the radiative efficient case (i.e. $>$\,1 per cent of the Eddington luminosity), an accretion disk 
forms which is optically thick, geometrically thin and surrounded by what is often referred to as a 
dusty `torus' (e.g. \citealt{Shakura_1973}). These sources fit into the unified scheme 
in which the different properties exhibited by AGN are mainly due to the obscuring 
effect of the dusty torus (see \citealt{Antonucci_1993} for a review). Observational 
evidence suggests that a small fraction of these ($\sim$15-20 per cent) are `radio-loud', 
meaning that they have ratios of radio-to-optical (e.g. 5\,GHz to B-band) flux above 10 
(\citealt{Kellerman_1989}). The resulting energy output 
from the accretion disk is radiated over a very broad range of frequencies and is expected to couple 
strongly to the gas inside the galaxy. Large-scale outflows and\,/\,or heating of gas and 
dust may be the result of such interactions (e.g. \citealt{Cattaneo_2009} and references 
therein). This implies a quenching of star-formation which then naturally sets up the tight 
relationship between bulge mass and black-hole mass. In addition, the radio jets of 
radiatively efficient radio-loud AGN may inject a large amount of energy into the hot phase of the 
intergalactic medium (IGM) fundamentally changing the halo environment, and possibly, 
the nature of subsequent AGN activity (e.g. \citealt{Rawlings_2004}). 
For radiatively inefficient AGN (i.e. $<$\,1 per cent of the Eddington luminosity),
accretion onto the black hole leads to very little radiated energy, making it 
difficult to identify such objects. However, those hosting jets may be identified 
at radio frequencies. These are thought to be fuelled by 
advection-dominated accretion flows (ADAFs) which are optically thin and 
geometrically thick (e.g. \citealt{Hardcastle_2007}). 

The radio-loud AGN activity of the most massive galaxies has been used in models to induce 
the sharp upper cutoff in the observed optical luminosity function of low redshift galaxies 
(e.g. \citealt{Croton_2006}). The fraction of galaxies showing radio-loud activity is a 
very strong function of stellar mass (e.g. \citealt{Best_2005b}; \citealt{Mauch_2007}), 
providing a natural selection function for solving the bright-end problem.
Feedback, in the form of radio jets, is used to deposit large amounts of energy into 
the halo, preventing the hot gas from cooling and providing fuel for more star formation 
(see \citealt{McNamara_2012} for a review on mechanical feedback from AGN). 
However, the details of this model remain unclear: what is the origin and nature
of the gas fuelling the AGN, how is the gas accreted onto the SMBH and by what mechanism
are the resulting relativistic jets coupled to the hot phase of the IGM?  
Much of the recent observational work on radio galaxies attempts to elucidate aspects 
of the feedback cycle. For example, it has been shown that the time averaged mechanical energy 
output from the radio jets of most early-type galaxies is sufficient to balance 
gas cooling from the hot phase of the IGM (\citealt{Best_2006}; \citealt{Best_2007}).
In addition, there is evidence for a tight correlation between the Bondi accretion 
rate (inferred from X-ray data) and the power emerging from these systems in 
relativistic jets (e.g. \citealt{Allen_2006}); this suggests that, in at least some massive 
galaxies, the fuelling mechanism is linked to the AGN activity. However, because 
radio jets are observed in a variety of galaxies, isolating which radio galaxies may be
involved in feedback-related quenching requires an understanding of the radio 
source population.

At low radio luminosities, radio galaxies are dominated by inefficiently accreting 
black holes with the central region contributing little from the X-ray to the far-infrared 
(e.g. \citealt{Hardcastle_2009} and references therein). These objects have been 
traditionally called low-excitation radio galaxies (LERGs) due to the lack of strong 
optical emission lines in the spectra of such sources (e.g.. \citealt{Hine_1979}; 
\citealt{Laing_1994}; \citealt{Jackson_1997}). Indeed, the results from \citet{Herbert_2011}
suggest that low luminosity radio galaxies occupy the same region of the fundamental 
plane as inactive galaxies.
At higher radio luminosities ($>$1$\times$10$^{26}$ W\,Hz$^{-1}$ at 1.4-GHz; 
\citealt{Best_2012}), the radio source population is dominated by high-excitation 
radio galaxies (HERGs). 
The optical spectra suggest accretion efficiencies that imply a fundamentally 
different type of AGN activity from that of LERGs, although the cause of this difference 
remains unclear. One proposal, put forward by \citet{Hardcastle_2007}, involves 
the origin of the fuelling gas. In this scenario, LERGs are fuelled by the hot phase 
of the IGM via Bondi accretion. This limits LERGs to hottest, densest and richest 
environments most easily provided by the halos of massive ellipticals (\citealt{Dekel_2006}; 
\citealt{Keres_2005}).
Conversely, HERGs are fuelled by cold gas transported to the 
central regions by what are, presumably, on-going gas-rich mergers. This is 
consistent with the finding that a large fraction of HERGs show evidence for 
disturbed morphology in the optical (\citealt{Ramos_Almeida_2011}) and
recent star formation in the ultraviolet (\citealt{Baldi_2008}). The nature of the fuelling
mechanism allows HERGs to occupy a much broader mass range than LERGs,
and in particular they are not expected to be restricted to the most massive galaxies.
Indeed, \citet{Best_2012} have shown that HERGs have systematically 
lower masses than LERGs and have 4000-\AA\, break strengths that suggest 
much higher levels of star-formation, in agreement with \citet{Herbert_2010}.

Most stars form in dense clouds of gas and dust. These stars produce copious 
amounts of ultraviolet (UV) radiation that is absorbed and re-radiated by the 
surrounding dust, heating it. The resulting spectrum peaks in the far-infrared 
(FIR)\,/\,sub-mm with temperatures ranging from 15-60\,K. In addition, dust in 
galaxies tends to be optically thin at FIR wavelengths allowing the dust masses 
and star-formation rates of galaxies to be estimated (e.g. \citealt{Kennicutt_1998,
Kennicutt_2009}). 
FIR\,/\,sub-mm studies of star-formation in samples of radio
galaxies have generally concentrated on high-redshift objects, in which
emission at long observed wavelengths (e.g. 850 $\mu$m, 1.2 mm)
corresponds to rest-frame wavelengths around the expected peak
of thermal dust emission (e.g. \citealt{Archibald_2001}; \citealt{Reuland_2004}). 
Working at these high redshifts with the available flux-limited samples 
in the radio necessarily restricted these studies to the most powerful 
radio-loud AGN. However, the availability of observations at shorter 
FIR\,/\,sub-mm wavelengths with the {\it Herschel Space Observatory} 
\citep{Pilbratt_2010} opens up the possibility of studies of very large 
populations of more nearby objects. The  {\it Herschel} Astrophysical 
Terahertz Large Area Survey (H-ATLAS: \citealt{Eales_2010}) has mapped 
550 deg$^{2}$ in 5 photometric bands (100, 160, 250, 350 \& 500$\mu$m)
with 5$\sigma$ sensitivities of 132, 121, 33.5, 37.7, 44.0\,mJy respectively 
(\citealt{Rigby_2011}). Using the Science demonstration data ($\sim$ 
14.4 deg$^{2}$), \citet{Hardcastle_2010} (hereafter H10) studied the 
FIR properties of radio galaxies out to redshift 0.85. This work sought to 
establish, on a statistical basis, whether radio-detected galaxies have different 
star-formation rates than their non radio-detected counterparts. H10 found no 
evidence for any difference, although the low source count ($\sim$ 200) 
made it impossible to investigate the FIR luminosity dependence on parameters 
such as galaxy mass and optical colours.

In this paper we investigate the FIR luminosity and dust temperature of radio-detected 
galaxies as a function of radio luminosity, redshift, radio-source size and radio spectral index.
There are a number of observational advantages when selecting AGN in the radio. 
Firstly, radio emission is unaffected by obscuring dust. Secondly, large scale surveys 
at radio frequencies, in combination with deep near-IR photometry, allow large numbers 
of sources to be identified across a wide redshift range. Thirdly, the size of the observed
radio jets can provide a rough indication of the age of the radio source and consequently, 
the AGN (e.g. \citealt{Kaiser_1997}). 
We present an analysis of all the radio-selected objects
identified with galaxies in the 134.7 deg$^{2}$ H-ATLAS Phase 1 field.
This represents an order of magnitude increase in survey area over H10 
and thus gives us much better statistics, which will allow us to investigate a 
wider range of host galaxy properties as a function of the FIR luminosity.
Our primary aim is to determine which parameters control the FIR 
luminosity output of radio galaxies and how. Such an understanding is important
if we are to fully understand how radio jets influence their hosts.
The paper is structured as follows. In Section\ \ref{CoG} we list the various
sources of data we have used in the following analysis, and 
describe the general sample selection. In Section\ \ref{radio-sample} we describe
the radio sample selection and the properties of these sources. In
Section\ \ref{far-IR-prop} we describe how the FIR luminosities and dust temperatures
of our sources were calculated followed by a discussion of the results in Section\ \ref{results}.
We then discuss the implications of our results and conclude in Section\ \ref{discussion}
and\ \ref{conclusions}, respectively.

Throughout the paper we use a concordance cosmology with 
$H_0 = 70$ km s$^{-1}$ Mpc$^{-1}$, $\Omega_{\rm m} = 0.3$ 
and $\Omega_\Lambda = 0.7$. All photometry is in Vega unless
specified otherwise.


\begin{figure*}
\begin{center}
\includegraphics[scale=0.6,clip,trim=0.6cm 0.1cm 0cm 0cm]{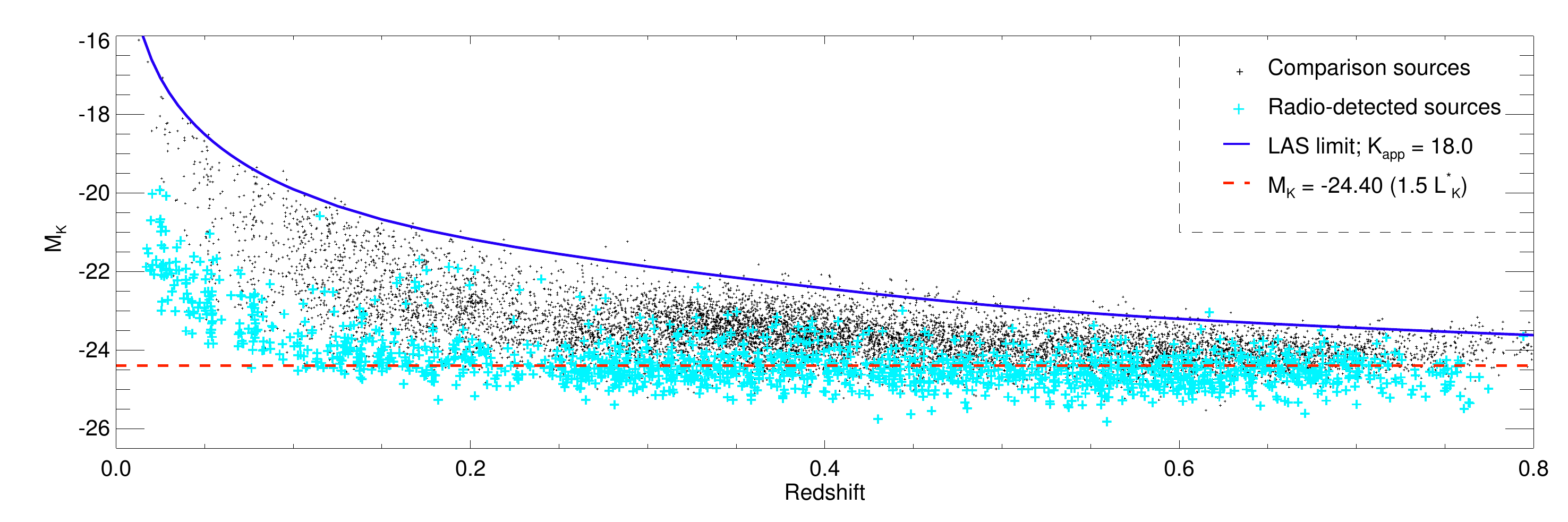}
\caption{Absolute $K$-band magnitude (M$_K$) versus redshift for galaxies
in the radio-detected and comparison samples. Blue crosses are the K-corrected, 
passive evolution corrected radio-detected sources. Black dots are the non radio-detected 
comparison sources. For clarity, we only plot 0.3 per cent of the comparison 
sources (randomly selected). The dashed red line, at -24.40, separates the 
sub- and super-1.5\,$L_{K}^{*}$ samples. The solid dark blue line is the depth 
of the LAS in the K-band ($K=18.0$).}
\label{kz}
\end{center}
\end{figure*}

\section{The data}
\label{CoG}

In this section we give an overview of the data used throughout this paper. 

\begin{enumerate}

\item Radio source catalogues and images from the NRAO VLA
Sky Survey (NVSS; \citealt{Condon_1998}) and Faint Images of 
the Radio Sky at Twenty-one centimetres (FIRST; \citealt{Becker_1995}) 
survey. These cover the entire H-ATLAS Phase 1 area (hereafter, 
simply Phase 1 area).
\item Radio source catalogues and images from the Giant Metrewave 
Radio Telescope at 330\,MHz (hereafter referred to as GMRT images 
and catalogues) covering $\sim$\,82 per cent of the Phase 1 area (Mauch 
et al. in preparation).
\item Point spread function (PSF) convolved, background subtracted
images of the Phase 1 fields at the wavelengths of 250, 350 and 
500\,$\mu$m, provided by the Spectral and Photometric Imaging 
REceiver (SPIRE) instrument on \emph{Herschel} \citep{Griffin_2010}.
The Phase 1 area consists of three equatorial strips each centred at 
9$^{h}$,  12$^{h}$, 14.5$^{h}$. The construction of these maps is 
described in detail by \citet{Pascale_2011}. 
\item Photodetector Array Camera and Spectrometer (PACS; 
\citealt{Poglitsch_2010}) maps at the wavelengths of 100 and
160\,$\mu$m. The reduction process for these maps is described
in detail by \citet{Ibar_2010}.
\item Catalogues and images from the United Kingdom Infrared 
Telescope Deep Sky Survey - Large Area Survey (UKIDSS-LAS, 
hereafter LAS; \citealt{Lawrence_2007}). The LAS covers $\sim$\,95.6 
per cent of the Phase 1 area. 
\item A set of photometric redshifts for galaxies detected by either 
the Sloan Digital Sky Survey Data Release 7 (hereafter, SDSS; 
\citealt{Abazajian_2009}) or LAS. These redshifts were generated
using the well known artificial neural network code ANNZ 
(\citealt{Collister_2004}). More details on the generation of
the photometric redshifts is described in \citet{Smith_2011} 
(hereafter, S11). This catalogue provides a redshift
estimate for every source detected in the optical and NIR
bands.
\item A catalogue of identifications between optically detected 
galaxies and the H-ATLAS data (Valiante et al. in preparation and
Hoyos et al. in preparation). For the identifications 
between the SDSS and the Phase 1 catalogue, a 
reliability $R$ is defined which is a measure of whether a single optical 
($r$-band) source dominates the observed FIR emission; they 
suggest that only sources with $R>0.8$ be used for this to be 
the case. Throughout the paper we consider all sources in the 
H-ATLAS catalogue, but distinguish in our analysis between `reliable' 
($R>0.8$) and unreliable identifications.
\item Redshifts from the Galaxy and Mass Assembly (GAMA) survey
(\citealt{Driver_2009,Driver_2011}). GAMA is a deep spectroscopic 
survey with limiting depths of $r_{\rm AB} < 19.8$ mag, $z < 18.2$ and $K_{\rm AB}
< 17.6$ over the Phase 1 area; details of the target selection and priorities 
are given by \citet{Baldry_2010}. 
The GAMA catalogue (SpecAllv14) for this area contains 131,921 new spectroscopic 
redshifts in addition to 10,351 redshifts from previous surveys in the area.

\end{enumerate}

We first filtered the galaxy catalogue so as to require a $K$-band detection 
in the LAS (K\,$<$\,18) and an r$'$-band detection ($r'<22$) in the SDSS. 
Next, we cross-matched the catalogue from part (vi) above with the H-ATLAS 
catalogue using TOPCAT 
\citep{Taylor_2005} on a 1$''$ best-match basis in order to identify which sources
had been detected at the 5$\sigma$ limit. In order to gain spectroscopic redshifts, we 
cross-matched the resulting catalogue with the GAMA catalogue on a 1$''$ 
best-match basis.
We selected spectroscopic redshifts 
(from GAMA and SDSS) in preference to photometric redshifts and filtered 
sources not in the redshift range $0.01<z<0.8$. If a photometric redshift was used, 
we required the error on the redshift to be $<$\,20 per cent of the redshift 
(see S11 for a discussion of the errors). Approximately 24 per cent of our 
sample have spectroscopic redshifts. 98 per cent of the
spectroscopic redshifts are below $z$\,$<$\,0.5. In order to reliably calculate the 
stellar mass (using the $K$-band) and the optical colour (g$'$-\,r$'$) we filter so as 
to exclude any sources that are point-like in either the LAS or the SDSS parent 
catalogues. K-corrected g$'$-\,r$'$ colours were then calculated using KCORRECT v4.2 
(\citealt{Blanton_2007}) for the whole sample. This process gave us a total 
of 324,222 galaxies in the Phase 1 area from which we separate the 
radio-detected galaxies in the next section.


\section{The radio-detected sample}
\label{radio-sample}

To produce the radio galaxy sample
we selected all catalogued NVSS sources in the Phase 1 field (7074 sources) above 
2.25\,mJy, the 5$\sigma$ limit for the NVSS catalogue. The good short-baseline coverage
of the NVSS data ensures that the NVSS flux densities are good estimates of the
true total flux density of our targets. Although the NVSS is only $\sim25$ per cent complete at 
2.25\,mJy (rising to $\sim100$ per cent at 3.5\,mJy), we retain this lower flux cut since
we expect no bias to be introduced from the incompleteness. We estimate that
$<0.1$ per cent (or $\sim$\,230 sources) of the comparison sample may be radio-loud 
(this is because the NVSS source extraction software does not extract 100 per cent of the
radio sources below 3.5\,mJy; see \citealt{Condon_1998}), but this 
is negligible given that there are over 320,000 comparison sources. In addition, none of 
our analysis is strongly dependent on the completeness of our sample. 

Next, we manually inspected all the NVSS sources in the field in order to
ensure that the correct host was identified for those radio sources which 
have complex structures (e.g. doubles). 
We did this by cross-matching to the LAS by overlaying
radio contours (both from FIRST and NVSS which have resolutions of $\sim$5$''$ 
and $\sim$45$''$ respectively) on LAS $K$-band images, accepting only sources which had an
association between the FIRST or, in relatively few cases (94), NVSS
radio images and a $K$-band object with the appearance of a galaxy. 
FIRST is used in preference to the NVSS for identifications because of its much
higher angular resolution, allowing less ambiguous identifications. Where a 
single compact radio component was present, the associated LAS source is
never more than 2.5$''$ from the FIRST position.
This process excludes some weak or diffuse NVSS sources where solid FIRST 
detections were not available and where the NVSS position is inadequate to allow an
identification with an LAS source. In addition, where NVSS sources
were found to be blends of two or more FIRST sources, we corrected
the NVSS flux density by scaling it by the ratio of the integrated FIRST flux
of the nearest source to the total integrated FIRST flux. As mentioned above,
the LAS covers only 95.6 per cent of the area of the Phase 1 field, so the choice
to use this as our reference catalogue slightly reduces our coverage
but does not affect the sample completeness in any way. In order to determine 
(as far as can be measured) the reliability of the cross-matching 
method, approximately twelve per cent (905) of the radio sources were cross matched 
for a second time another individual. Encouragingly, the disagreement rate was 
$<1$ per cent. This process gave us a total of 3182 radio sources with LAS 
identifications.

\begin{figure*}
\begin{center}
\includegraphics[scale=0.45,clip,trim=0cm 0cm 0cm 0cm]{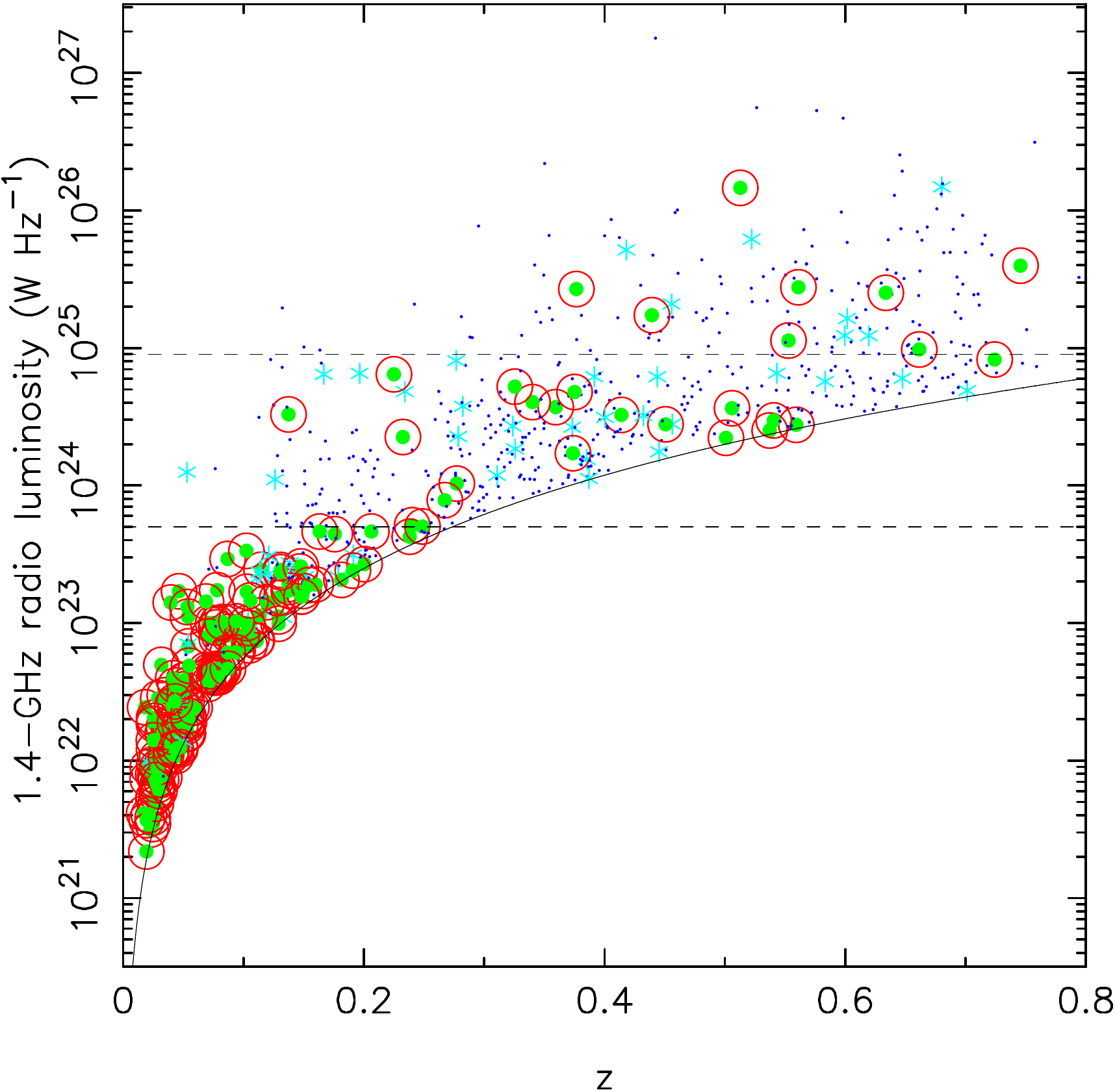}
\includegraphics[scale=0.45,clip,trim=-2cm 0cm 0cm 0cm]{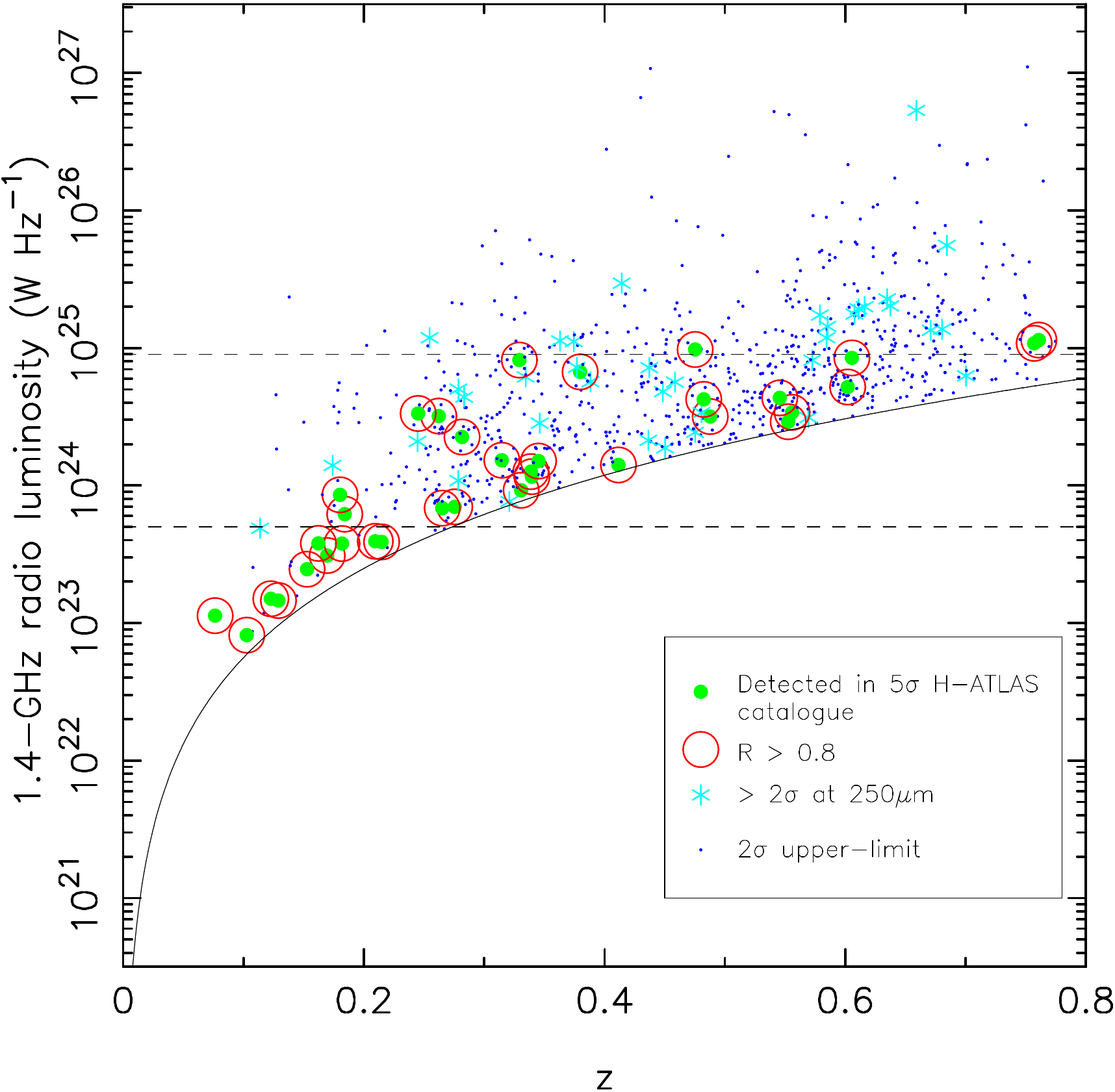}
\caption{Radio luminosity of the radio-selected sample as a function of
  redshift split into the sub-1.5\,$L_{K}^{*}$ sample (left) and super-1.5\,$L_{K}^{*}$ 
  (right). Sources nominally detected by {\it Herschel} (i.e. sources identified with 
  H-ATLAS objects) are marked as filled green circles. If the association with the 
  LAS galaxy is deemed `reliable', the object is also marked with a red open circle. 
  Any source not identified with H-ATLAS object in the catalogue, but is detected 
  down to the $2\sigma$ level (see Section\ \ref{FIRintro} for how we define the $2\sigma$
  level) in the 250-$\mu$m images, is shown as a light blue 
  star for illustrative purposes only. Non-detections are marked with blue dots. The solid 
  line corresponds to the 
  nominal 5$\sigma$ flux cutoff in the NVSS of 2.25\,mJy. Dashed horizontal lines 
  show radio luminosities corresponding to (from bottom to top) the point at which 
  the starburst radio luminosity function cuts off steeply (\citealt{Sadler_2002}) and 
  the expected luminosity for a maximal (4000 $M_\odot$ year$^{-1}$) starburst.}
\label{lz}
\end{center}
\end{figure*}

Next, we cross-correlated this radio catalogue with our base catalogue 
(see previous section) using the LAS positions on a 1$''$ best match basis. 
This process gave us a total of 1599 radio-detected sources in the Phase 1 
area, of which 1110, or 69 per cent, had spectroscopic redshifts. Approximately 
80 per cent of the spectroscopic redshifts are below $z=0.5$. The SDSS $r'$-band 
limit accounts for the reduction of LAS identified radio sources from 3182 
to 1599; $\sim$\,76 per cent of the dropouts have $K$\,$>$\,17.0. 

Next, we measured the size of all the cross-matched radio sources. If the source was 
compact in FIRST, we used the de-convolved major axis from the FIRST catalogue
as the projected size of the radio source. The 1$\sigma$ source size error
is given by:
\begin{equation}
\mathrm{Sigma} =  10\,''\,\times\,\left(\frac{1}{SNR} + \frac{1}{75}\right)
\end{equation}
where SNR is the signal-to-noise ratio of the FIRST source (\citealt{Becker_1995}).
If the de-convolved source size was less than 2$\sigma$ (equivalent to the 95 per cent 
contour) we took a 2$\sigma$ upper limit; this was 
the case for $\sim$\,13 per cent of the sources. The sizes of more complicated structures such as 
double-lobed radio sources were measured by calculating the projected distance 
between the peaks of the lobes. Where there was evidence for multiple lobes,
the distance between the two furthest lobes from the host was recorded. There were 89 
objects in which the NVSS emission was resolved out in FIRST. Such sources were not
assigned sizes and were thus excluded from the analysis involving radio source sizes.

In order to calculate spectral indices\footnote{Where the spectral index
$\alpha$ is defined in the sense that $S \propto \nu^{-\alpha}$}, 
we cross-matched the filtered 1.4-GHz 
radio catalogue with 330\,MHz GMRT radio catalogues and images, whose
depth range from 3-10\,mJy (Mauch et al. in preparation). Each match 
was checked manually by overlaying GMRT radio contours on LAS K-band 
images. A total of 726 sources were successfully matched. Upper limits 
on the spectral indices were calculated for the remaining sources (654) 
by substituting the GMRT flux by five times the local RMS. A further 219 
radio sources do not have spectral indices or limits due to limited GMRT 
coverage of the Phase 1 area. This catalogue, which is a subset of the 
main radio catalogue described above, is only used in the analysis 
involving spectral indices.

The above processes gave us a catalogue which is flux-limited in the radio
(by virtue of the original selection from the NVSS) and also 
magnitude-limited in the optical (we require a $K$-band identification
and also require $r'<22$). The result is that we have few (71) sources 
above $z>0.7$. The catalogue is also likely to be strongly biased 
against radio-detected quasars since we have excluded sources that appear 
point-like from our galaxy catalogue. This has the desirable effect that the 
measured FIR luminosities will tend not to be strongly affected by beaming of
the radio jet and that any contamination of the fluxes measured at {\it Herschel} 
wavelengths by non-thermal emission might be expected to be limited 
(cf. the results of \citealt{Hes_1995}). 
Hereafter, the sources not identified with radio emission are referred 
to as comparison or non radio-detected sources while the sources identified 
with radio emission are referred to as radio-detected sources.

\subsection{K-band luminosity cuts}

In order to account for any stellar mass dependent relations, we calculate the 
$K$-band luminosity. The $K$-band luminosity is dominated by old stellar light 
(e.g. \citealt{Best_1998, C.Simpson_1999}) and thus scales well with stellar 
mass. In order to enable us to compare radio galaxies with their mass equivalent 
non radio-detected counterparts, we use the absolute $K$-band magnitude.

Fig.\ \ref{kz} shows the absolute $K$ band magnitude versus redshift.
Here we calculate the K-corrected and passive evolution corrected
magnitude at $K$ using a GALAXEV starburst spectral energy distribution 
(see \citealt{Bruzual_2003}) at $z=4$, with solar metallicity, allowing it to 
exponentially decay with $\tau=1$\,Gyr.
We apply this correction to both the radio sources and the comparison galaxies. 
We adopt a value of $-23.96$ for M$^{*}_{K}$ (\citealt{A.Smith_2009}; value 
calculated for our adopted cosmology) from which we calculate $L_{K}^{*}$ to 
be 7.87\,$\times$\,10$^{10}$\,$L_{\astrosun}$.
We separate the radio-selected sample into two magnitude bins, defined by 
M$_{K}$\,$>$\,-24.4 (sub-1.5\,$L_{K}^{*}$) and M$_{K}<-24.4$ 
(super-1.5\,$L_{K}^{*}$). For illustrative purposes only, we calculate the stellar 
mass boundary assuming a mass-to-light ratio of 0.6 (\citealt{Kauffmann_1998}) 
and find it to be roughly 7.1$\times$10$^{10}$\,M$_{\astrosun}$. The lower-luminosity 
sample contains 770 radio sources while the higher-luminosity sample 
contains 829 radio sources.

\subsection{Radio luminosities}
\label{rad-lum}

Fig.\ \ref{lz} shows 1.4-GHz radio luminosity versus redshift for our two $K$-band 
luminosity separated radio samples. Since less than half of the radio 
sources at 1.4-GHz are detected at 330\,MHz, we adopt $\alpha =
0.8$ when calculating the 1.4-GHz radio luminosity of \emph{all} the radio 
sources, where $S \propto \nu^{-\alpha}$ for the K-correction in the radio luminosity 
calculations; $\alpha = 0.8$ is a typical observed value for low-frequency selected 
objects, and we expect 
that the selection against point-like optical objects will tend to select against
flat-spectrum radio sources. Indeed, the subset of sources with GMRT detections 
suggest that only $\sim$\,2 per cent of our sources have $\alpha < 0$ and
that $\alpha = 0.8$ is a reasonable value for our sample.\\

It is clear from Fig.\ \ref{lz} that we probe a wide range of radio luminosities. At the low-luminosity, 
low redshift end ($z<0.2$), we expect from existing analysis of the local 1.4-GHz 
luminosity function (e.g. \citealt{Sadler_2002}; \citealt{Mauch_2007}) that the 
population will be dominated by star-forming galaxies rather than radio
AGN, although a few AGN may still be present. The starburst luminosity function 
cuts off steeply above $5\times10^{23}$\,W\,Hz$^{-1}$ at 1.4-GHz, so we expect 
that most of the objects above this luminosity will be radio AGN. Thus, in our 
sub-1.5\,$L_{K}^{*}$ radio sample there exist two populations; the low redshift, 
low luminosity starburst population and the higher redshift, higher luminosity 
radio AGN population. 

The super-1.5\,$L_{K}^{*}$ radio sample contains fewer sources below 
$5\times10^{23}$\,W\,Hz$^{-1}$ and are thus almost exclusively composed
of radio AGN. As mentioned in Section\ \ref{intro}, 
\citet{Best_2012} calculated the HERG\,/\,LERG luminosity functions for the local 
Universe and found that HERGs begin to dominate above $10^{26}$\,W\,Hz$^{-1}$. 
Thus, our samples will be dominated by LERGs, although significant numbers of 
HERGs may be present, particularly at lower masses.\\


\section{The far-IR properties of the sample}
\label{far-IR-prop}

\subsection{\emph{Herschel} flux density measurements}
\label{FIRintro}

In this section we describe the properties of the sample in the
FIR. Throughout this section, FIR flux densities in the SPIRE 
bands are measured directly from the background-subtracted, 
PSF-convolved H-ATLAS Phase 1 images described in Section 
\ref{CoG}, taking the best estimate of the flux density to be the 
value in the image at the pixel corresponding most closely to the 
LAS position of our targets, with errors estimated from the corresponding 
position in the noise map. As discussed by \citet{Pascale_2010},
PSF-convolved maps provide the maximum-likelihood estimate 
for the flux density of a single isolated point source at a given 
position in the presence of thermal noise; this remains a reasonable 
approximation if the correlations between the positions of multiple 
sources are small, as we expect in real data due to physical clustering 
of objects. 
We also extracted PACS flux densities 
and corresponding errors from the images at 100 and 160\,$\mu$m
using circular apertures appropriate for the PACS beam (respectively 
15.0 and 22.5 arcsec) and using the appropriate aperture corrections, 
which take account of whether any map-pixels have been masked during the 
process of high-pass filtering of the bolometer timelines. We add 
an estimated absolute flux calibration uncertainty of 10 per cent (PACS)
and 7 per cent (SPIRE) in quadrature to the errors measured from
the maps for the purposes of fitting and stacking, as recommended
in H-ATLAS documentation.
To account for bright, extended objects, we use the flux densities in the 
Phase 1 5$\sigma$ catalogue, where available, in preference to our measured 
flux densities, even though our measured flux densities correlate well with the 
catalogued fluxes. To correct for confusion in the SPIRE maps, we 
subtract the mean flux density level of the whole PSF-convolved map from the 
flux density measurements of each source. This ensures that the mean of a randomly 
selected sample will be zero within the uncertainties.

Approximately 13.6 per cent of the radio sample (218 sources, which includes 
potential star-forming galaxies) is detected at the 5$\sigma$ limit demanded by 
the H-ATLAS source catalogue. This is broadly similar to the detection fraction 
found in H10. In order to give the reader a feel for the H-ATLAS $S/N$ distribution 
of the radio sample, we define a 2$\sigma$ rms value. Since the images at 250, 350 
and 500\,$\mu$m are seriously affected by confusion (as opposed to the PACS 
maps, which are not), we cannot use the 2$\sigma$ limit defined by simple Gaussian 
statistics (i.e twice the RMS). Instead, we follow \citet{Hardcastle_2013} (hereafter, 
H13) and randomly sample a large number of points in the 250, 350 and 500\,$\mu$m 
maps, selecting the flux value below which 97.7 per cent of the random fluxes lie as our 
2$\sigma$ limit. This gives flux limits of 24.6, 26.5 and 25.6\,mJy for the 250, 350 and 
500-mm SPIRE maps respectively. Therefore, sources which have a $S/N>2$ at 250, 
350 and 500\,$\mu$m make up 18.9, 16.6 and 12.6 per cent of the sample, respectively.

We note that whether a source is `detected' or not plays no part in the following analysis, 
since we are interested in the sample properties taken as a whole. However, in some 
of the following plots the FIR luminosity of sources above 2$\sigma$ are shown 
as individual points for illustrative purposes only.

\begin{figure*}
\includegraphics[scale=0.4,clip,trim=0cm 0.5cm 0cm 0cm]{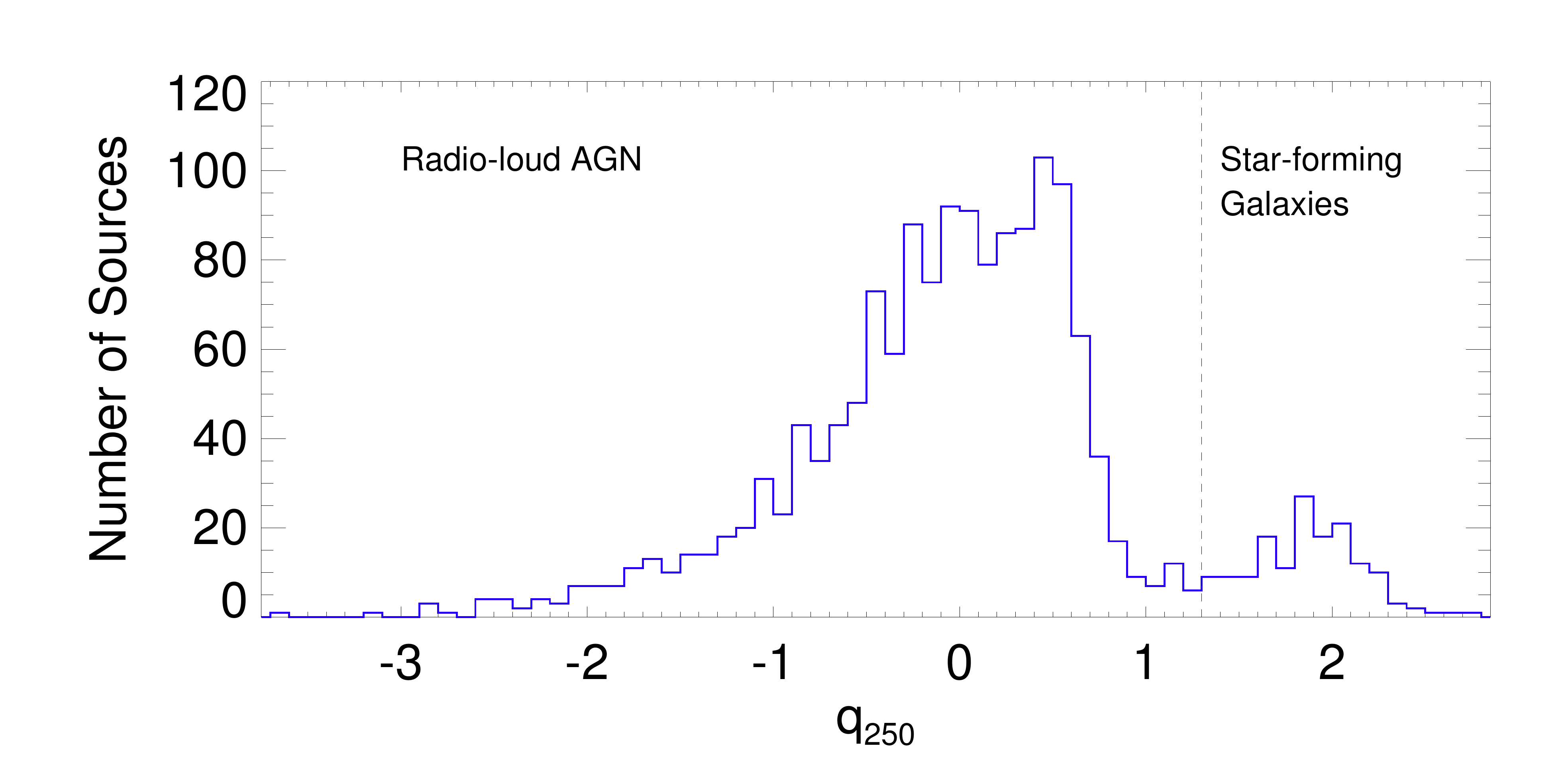}
\caption{Histogram of $q_{250}$ values, defined in Eq.\,2. We remove all sources
with $q_{250}>1.3$, in order to ensure no star-forming objects are in our radio
AGN sample. We note that many of the $q_{250}$ values are upper-limits, particularly 
below $q_{250}<1.3$.}
\label{q250}
\end{figure*}

\subsection{250$\mu$m luminosity calculations}
\label{l250}

In H10, we calculated the integrated IR-luminosity of individual 
sources by assuming a spectral energy distribution (SED) with 
$T=26$\,K and $\beta$\,=\,1.5 following \citet{Dye_2010}.
\footnote{H10 contained an error in the calculated K-correction which 
increased the estimated luminosities at high redshift. However, because 
our conclusions in that paper were not dependent on the \emph{absolute} 
luminosities, but rather the \emph{relative} luminosities, no science 
conclusions were affected.} 
In this paper, we follow a similar method to that of H10 with some key 
differences. Firstly, we assume $T=20$\,K and $\beta$\,=\,1.8, 
best-fitting values derived for a similarly selected sample of radio galaxies 
by H13. Secondly, we calculate the FIR luminosity at 250\,$\mu$m, rather 
than the integrated IR luminosity. This means that the isothermal model,
and hence the assumed temperature only act as a K-correction factor in the
determination of the 250\,$\mu$m luminosity ($L_{250}$) of each source. Further, we note
that at low redshifts, the calculated K-correction should have almost no effect
on the 250\,$\mu$m luminosity. The disadvantage of switching to the 250\,$\mu$m 
luminosity is that we are no longer able to estimate the 
SFR. However, the relations which convert the integrated IR luminosity 
to a SFR are calibrated for star-forming galaxies (e.g. \citealt{Kennicutt_1998}) 
and thus may not be applicable to the temperature and luminosity range 
radio galaxies occupy in any case.

The \emph{Herschel} SPIRE PSF has a FWHM of 
18$''$ at 250\,$\mu$m. This translates to linear sizes of $\sim$135\,kpc
at $z=0.8$, the redshift of our most distant objects. Therefore, the 
corresponding FIR luminosities and temperatures we measure apply not 
only to the host of the radio source but also to its immediate environment; 
this may include quiescent dust or star-formation in a companion or 
merging galaxy.

\subsection{Removing star-forming galaxies}
\label{Sec:SF}

In order to remove star-forming galaxies from our sample, we calculate $q_{250}$
as follows,
\begin{equation}
q_{250}=\log_{10}(L_{\rm 250}/L_{1.4})
\end{equation}
where $L_{250}$ is the K-corrected luminosity at 250\,$\mu$m. This was 
done for the entire radio-detected sample. A histogram of the results is shown in Fig.\ \ref{q250}. 
The far-infrared radio correlation (FIRC) has been shown to hold at 250\,$\mu$m 
(\citealt{Jarvis_2010}).
The expected dispersion for sources on the far-infrared radio correlation (FIRC) is 
$1.4<q_{250}<2.1$ (\citealt{Jarvis_2010}). Using Fig.\ \ref{q250}, we choose to remove 
all sources with $q_{250}>1.3$ or where the upper limit on $q_{250}$ is $>1.3$, ensuring 
that no (or a negligible few) star-forming objects are in our subsequent radio AGN sample. 
This ensures that any source with the potential to be a star-forming object has 
been removed. Therefore, beyond this point star-forming objects play no further part in the 
catalogues\,/\,analysis unless explicitly stated.

\begin{table*}
\scriptsize
\centering
\caption{Mean bin flux densities and K-S probabilities that the {\it Herschel} fluxes 
of radio-detected objects in redshift bins are drawn from the background distribution, 
as a function of wavelength. Low probabilities (below 1 per cent) imply significant 
differences between the bin being considered and the distribution of flux densities 
measured from randomly selected positions in the sky, as described in the text. 
Sources on the FIRC are not included.}
\begin{tabular}{llrrrrrrrrrrr}
\hline
Radio&$z$ range&Objects&\multicolumn{5}{c}{Mean bin flux density (mJy)}&\multicolumn{5}{c}{K-S probability (per cent)}\\
Sample&&in bin& \multicolumn{3}{c}{SPIRE bands}&\multicolumn{2}{c}{PACS bands}& \multicolumn{3}{c}{SPIRE bands}&\multicolumn{2}{c}{PACS bands}\\
&&&250 $\mu$m&350 $\mu$m&500 $\mu$m&100 $\mu$m&160 $\mu$m&250 $\mu$m&350 $\mu$m&500 $\mu$m&100 $\mu$m&160 $\mu$m\\
\hline
$<$1.5\,$L_{K}^{*}$&0.00 -- 0.15 &  92  & $25.0 \pm 0.8$  & $9.8 \pm 0.8$  & $6.9 \pm 0.9$  & $36.7 \pm 3.2$  & $34.8 \pm 3.9$ & $<10^{-3}$ & $<10^{-3}$ &0.02 & 0.04 & 0.3 \\
&0.15 -- 0.25 &  90  & $9.0 \pm 0.7$  & $3.1 \pm 0.8$  & $2.9 \pm 0.9$  & $12.8 \pm 3.0$  & $21.0 \pm 3.8$ & $<10^{-3}$ & 0.09 & 2.2 & 16.5 & 0.6 \\
&0.25 -- 0.40 & 182  & $7.0 \pm 0.5$  & $1.3 \pm 0.5$  & $2.4 \pm 0.6$  & $12.1 \pm 2.2$  & $6.4 \pm 2.8$ & $<10^{-3}$ & 1.4 & 5.8 & 1.5 & 82.7 \\
&0.40 -- 0.50 &  78  & $7.5 \pm 0.7$  & $2.8 \pm 0.8$  & $5.0 \pm 1.0$  & $1.6 \pm 3.4$  & $21.6 \pm 4.2$ & $<10^{-3}$ & 0.02 & 0.006 & 92.4 & 0.03 \\
&0.50 -- 0.60 &  75  & $9.1 \pm 0.7$  & $3.9 \pm 0.8$  & $4.9 \pm 1.0$  & $9.5 \pm 3.6$  & $19.8 \pm 4.4$ & $<10^{-3}$ & 0.2 & 0.5 & 42.1 & 12.2 \\
&0.60 -- 0.80 & 113  & $5.7 \pm 0.6$  & $2.2 \pm 0.7$  & $4.5 \pm 0.8$  & $1.7 \pm 3.0$  & $12.9 \pm 3.8$ & $<10^{-3}$ & 0.002 & 0.001 & 44.0 & 8.0 \\
\hline
$>$1.5\,$L_{K}^{*}$ &0.00 -- 0.15 &  11  & $6.7 \pm 1.9$  & $4.6 \pm 2.2$  & $3.8 \pm 2.5$  & $3.2 \pm 9.0$  & $12.7 \pm 11.0$ & 4.3 & 9.0 & 22.4 & 70.7 & 88.9 \\
&0.15 -- 0.25 &  40  & $9.8 \pm 1.0$  & $3.2 \pm 1.1$  & $3.0 \pm 1.3$  & $25.4 \pm 5.0$  & $34.4 \pm 6.3$ & 1.2 & 2.4 & 29.6 & 8.8 & 0.9 \\
&0.25 -- 0.40 & 236  & $6.4 \pm 0.4$  & $2.3 \pm 0.5$  & $3.1 \pm 0.5$  & $6.6 \pm 2.0$  & $10.8 \pm 2.5$ & $<10^{-3}$ & $<10^{-3}$ & 0.2 & 37.0 & 6.1 \\
&0.40 -- 0.50 & 140  & $2.7 \pm 0.5$  & $-0.3 \pm 0.6$  & $1.1 \pm 0.7$  & $-1.5 \pm 2.5$  & $0.9 \pm 3.1$ & 0.04 & 15.5 & 40.4 & 83.2 & 42.0 \\
&0.50 -- 0.60 & 178  & $2.2 \pm 0.5$  & $-0.0 \pm 0.5$  & $1.3 \pm 0.6$  & $-4.6 \pm 2.3$  & $4.3 \pm 2.9$ & 0.2 & 1.1 & 4.7 & 41.8 & 85.2 \\
&0.60 -- 0.80 & 210  & $4.1 \pm 0.4$  & $1.3 \pm 0.5$  & $3.3 \pm 0.6$  & $3.1 \pm 2.1$  & $5.2 \pm 2.7$ & $<10^{-3}$ & 0.003 & 0.1 & 4.4 & 12.2 \\
\hline
\end{tabular}
\label{ks-z}
\end{table*}

\begin{table*}
\scriptsize
\caption{Mean bin flux densities and K-S probabilities that the {\it Herschel} fluxes 
of radio-detected objects in radio luminosity bins are drawn from the background distribution, 
as a function of wavelength. Low probabilities (below 1 per cent) imply significant 
differences between the bin being considered and the distribution of flux densities 
measured from randomly selected positions in the sky, as described in the text. 
Sources on the FIRC are not included.}
\begin{tabular}{llrrrrrrrrrrr}
\hline
Radio&Range in&Objects&\multicolumn{5}{c}{Mean bin flux density (mJy)}&\multicolumn{5}{c}{K-S probability (per cent)}\\
Sample&log$_{10}$($L_{1.4}$)&in bin& \multicolumn{3}{c}{SPIRE bands}&\multicolumn{2}{c}{PACS bands}& \multicolumn{3}{c}{SPIRE bands}&\multicolumn{2}{c}{PACS bands}\\
&&&250 $\mu$m&350 $\mu$m&500 $\mu$m&100 $\mu$m&160 $\mu$m&250 $\mu$m&350 $\mu$m&500 $\mu$m&100 $\mu$m&160 $\mu$m\\
\hline
$<$1.5\,$L_{K}^{*}$&21.0 -- 23.7  & 114  & $22.3 \pm 0.7$  & $8.1 \pm 0.7$  & $5.1 \pm 0.8$  & $34.0 \pm 2.8$  & $30.1 \pm 3.4$ & $<10^{-3}$ & $<10^{-3}$ & 0.07 & 0.005 & 0.10 \\
&23.7 -- 24.3  & 140  & $4.3 \pm 0.5$  & $-0.7 \pm 0.6$  & $1.1 \pm 0.7$  & $7.8 \pm 2.5$  & $9.0 \pm 3.2$ & $<10^{-3}$ & 17.4 & 41.9 & 1.5 & 51.3 \\
&24.3 -- 24.6  & 115  & $9.2 \pm 0.6$  & $2.2 \pm 0.7$  & $3.3 \pm 0.8$  & $9.2 \pm 2.8$  & $21.5 \pm 3.4$ & $<10^{-3}$ & 0.8 & 0.6 & 56.4 & 3.0 \\
&24.6 -- 25.0  & 116  & $8.8 \pm 0.6$  & $5.1 \pm 0.7$  & $6.4 \pm 0.8$  & $8.0 \pm 2.9$  & $13.6 \pm 3.7$ & $<10^{-3}$ & $<10^{-3}$ & 0.002 & 38.4 & 20.5 \\
&25.0 -- 25.6  &  99  & $7.3 \pm 0.6$  & $2.9 \pm 0.7$  & $3.9 \pm 0.8$  & $3.7 \pm 3.1$  & $15.5 \pm 3.8$ & $<10^{-3}$ & 0.2 & 0.007 & 36.9 & 4.7 \\
&25.6 -- 27.3  &  46  & $7.7 \pm 1.0$  & $4.9 \pm 1.1$  & $7.7 \pm 1.3$  & $9.2 \pm 4.4$  & $13.4 \pm 5.4$ & 0.003 & 0.02 & 0.05 & 5.2 & 18.2 \\
\hline
$>$1.5\,$L_{K}^{*}$&21.0 -- 23.7  &  13  & $12.6 \pm 1.8$  & $6.4 \pm 2.0$  & $3.9 \pm 2.3$  & $7.3 \pm 8.4$  & $19.4 \pm 10.2$ & 1.2 & 8.0 & 7.5 & 59.3 & 25.9 \\
&23.7 -- 24.3  & 128  & $5.2 \pm 0.6$  & $1.2 \pm 0.6$  & $1.6 \pm 0.7$  & $6.8 \pm 2.8$  & $12.4 \pm 3.4$ & 0.002 & 0.3 & 13.2 & 8.4 & 5.9 \\
&24.3 -- 24.6  & 159  & $5.1 \pm 0.5$  & $3.0 \pm 0.6$  & $4.9 \pm 0.7$  & $1.6 \pm 2.4$  & $9.2 \pm 3.0$ & 0.009 & $<10^{-3}$ & 0.03 & 67.8 & 16.1 \\
&24.6 -- 25.0  & 268  & $3.2 \pm 0.4$  & $-0.3 \pm 0.4$  & $0.6 \pm 0.5$  & $3.0 \pm 1.9$  & $6.0 \pm 2.4$ & 0.006 & 0.9 & 34.0 & 56.0 & 51.2 \\
&25.0 -- 25.6  & 183  & $4.5 \pm 0.5$  & $1.5 \pm 0.5$  & $3.7 \pm 0.6$  & $-1.3 \pm 2.2$  & $5.3 \pm 2.8$ & $<10^{-3}$ & 0.06 & 0.2 & 56.6 & 1.3 \\
&25.6 -- 27.3  &  64  & $3.9 \pm 0.8$  & $0.8 \pm 0.9$  & $1.6 \pm 1.0$  & $7.1 \pm 3.5$  & $2.5 \pm 4.6$ & 0.008 & 0.2 & 36.3 & 0.9 & 57.0 \\
\hline
\end{tabular}
\label{ks-l}
\end{table*}

\subsection{Stacking analysis}
\label{stackAn}

Since the vast majority of our radio AGN ($\sim95$ per cent) 
are undetected at the 5$\sigma$ limit of the Phase 1 catalogue, we 
need to use statistical methods to calculate the properties of the source 
population. We elected to stack the sample using the 100, 160, 250, 350 
and 500\,$\mu$m H-ATLAS maps in bins of redshift and radio luminosity 
which range from $0.01<z<0.8$ and $21.0<{\rm log}_{10}$($L_{1.4}$)\,$<27.3$.
We use the same bin sizes, in both radio luminosity and redshift, across 
both $K$-band luminosity separated samples in order to facilitate 
comparisons. Individual bin sizes were chosen so as to keep as many bins 
distinguished from the background as possible.

To establish quantitatively whether sources in the bins were
significantly detected, we measured flux densities from 100,000
randomly chosen positions in the field; a Kolmogorov-Smirnov (K-S)
test could then be used to see whether the sources from our sample
were consistent with being drawn from a population defined by the
random positions. Using a K-S test rather than simply considering the
calculated uncertainties on the measured fluxes allows us to account
for the non-Gaussian nature of the noise as a result of confusion.
If the fluxes in any given bin are significantly distinguished from the 
background on a K-S test and have a mean flux density higher than
that of the underlying noise (which should be zero, see 
Section\ \ref{FIRintro}), we can calculate the \emph{stacked luminosity}
using all the objects in that bin in the knowledge that the resulting value 
is not simply noise.

The results from this analysis are shown in Tables\ \ref{ks-z} and\ \ref{ks-l}. 
Both tables show the results for the $K$-band luminosity separated radio 
samples. Sources on the FIRC have been removed. 
As is evident from Tables\ \ref{ks-z} and\ \ref{ks-l}, most of the SPIRE bins 
are detected at the 95 per cent confidence level (i.e. there is a $<5$ per 
cent chance that the flux distribution has been randomly drawn from the 
same underlying distribution as the background). However, the 
PACS bins are mostly indistinguishable from the background. This is the 
result of the lower sensitivity of the PACS data. It is also clear that the K-S 
probabilities are almost always lower at 250\,$\mu$m than at any other band, 
with all of the bins at 250\,$\mu$m detected above the 99 per cent confidence 
level or better, except for the lowest redshift bin of the super-1.5\,$L^{*}_{K}$ 
sample. The higher sensitivity and smaller beam size at 250\,$\mu$m are likely 
explanations for this. Indeed, if one excludes those radio sources which are also 
in the 5$\sigma$ H-ATLAS catalogue from the K-S test, all the SPIRE bins 
are detected at the 96 per cent level or better except the super-$L^{*}_{K}$ 
$0.15<z<0.25$ bin, which is detected at the 83.8 per cent level. This provides 
further justification for calculating the rest-frame luminosity 
at 250\,$\mu$m. 
The KS probabilities produced by the comparison sources are not shown, 
but all the bins at 250\,$\mu$m were detected above the 99 per cent confidence 
level or better. 

In order to calculate stacked luminosities, we calculate the weighted average of 
$L_{250}$ for each radio luminosity or redshift bin. Individual sources are weighted 
using the reciprocal of the square of the error extracted from the 250\,$\mu$m noise 
map. In order to calculate errors on the stacked luminosity, we use the bootstrap 
method with replacement. This involves randomly selecting a subsample from the
relevant bin, while ensuring that the total number of sources in that bin remains
constant. This implies that the same source may be selected more than once and 
others omitted completely. From this new sample, a weighted $L_{250}$ value is calculated 
and stored. We repeat the process 300 times in order to get a distribution of weighted $L_{250}$
values. From this distribution, we select the 16$^{th}$ and 84$^{th}$ percentiles 
as our lower and upper limits respectively.
The advantage of bootstrapping is that no assumption is 
made on the shape of the luminosity distribution; instead, we use the distribution of 
the sample to calculate 1$\sigma$ errors. This automatically takes all sources of 
intrinsic dispersion, such as redshift errors and flux errors, into account. The errors 
plotted in all subsequent FIR luminosity plots are the 1$\sigma$ uncertainties derived
from bootstrapping.

\begin{figure*}
\includegraphics[scale=0.6,clip,trim=0cm -0.2cm -0.5cm 0cm]{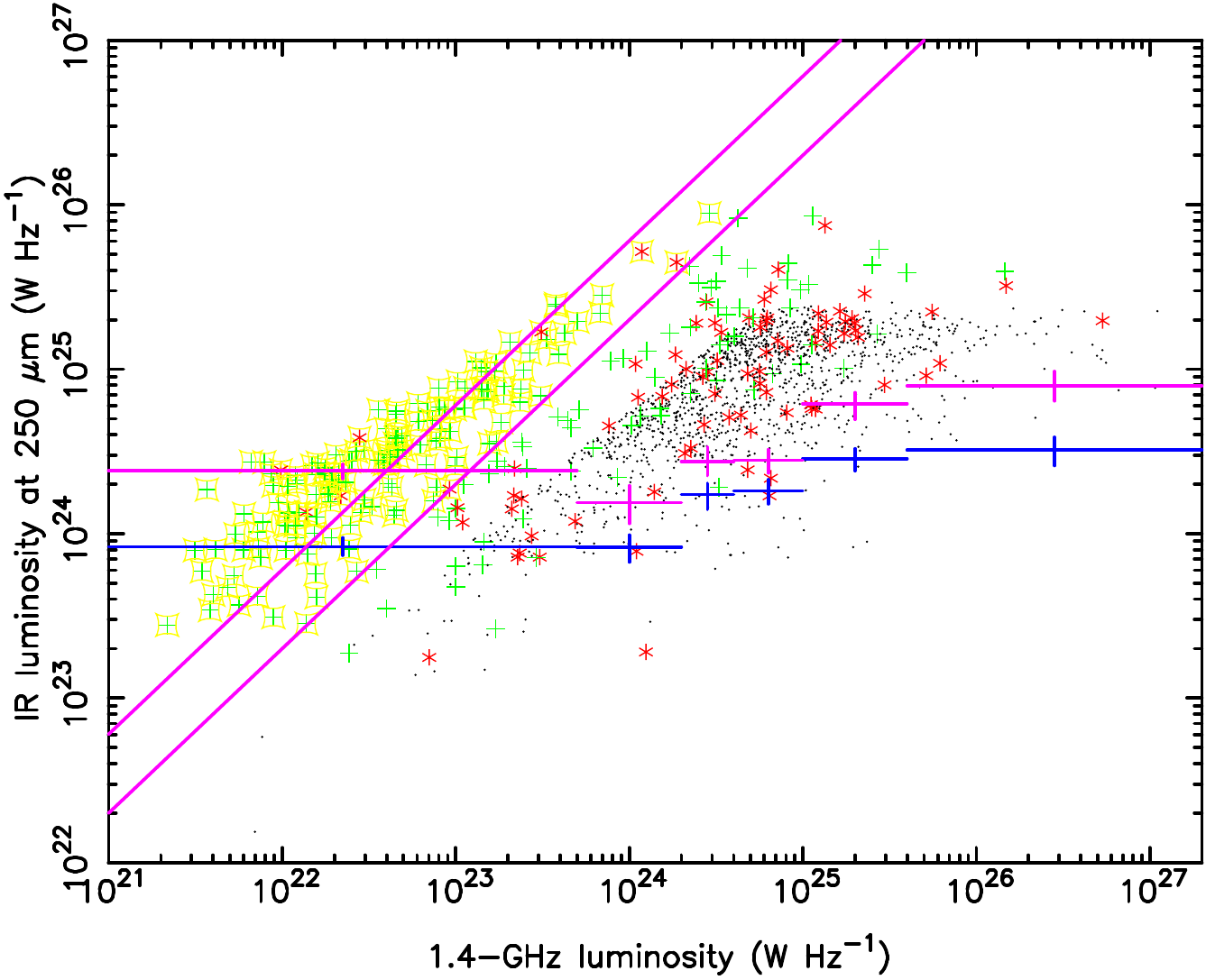}
\includegraphics[scale=0.6,clip,trim=0cm -0.2cm 0cm 0cm]{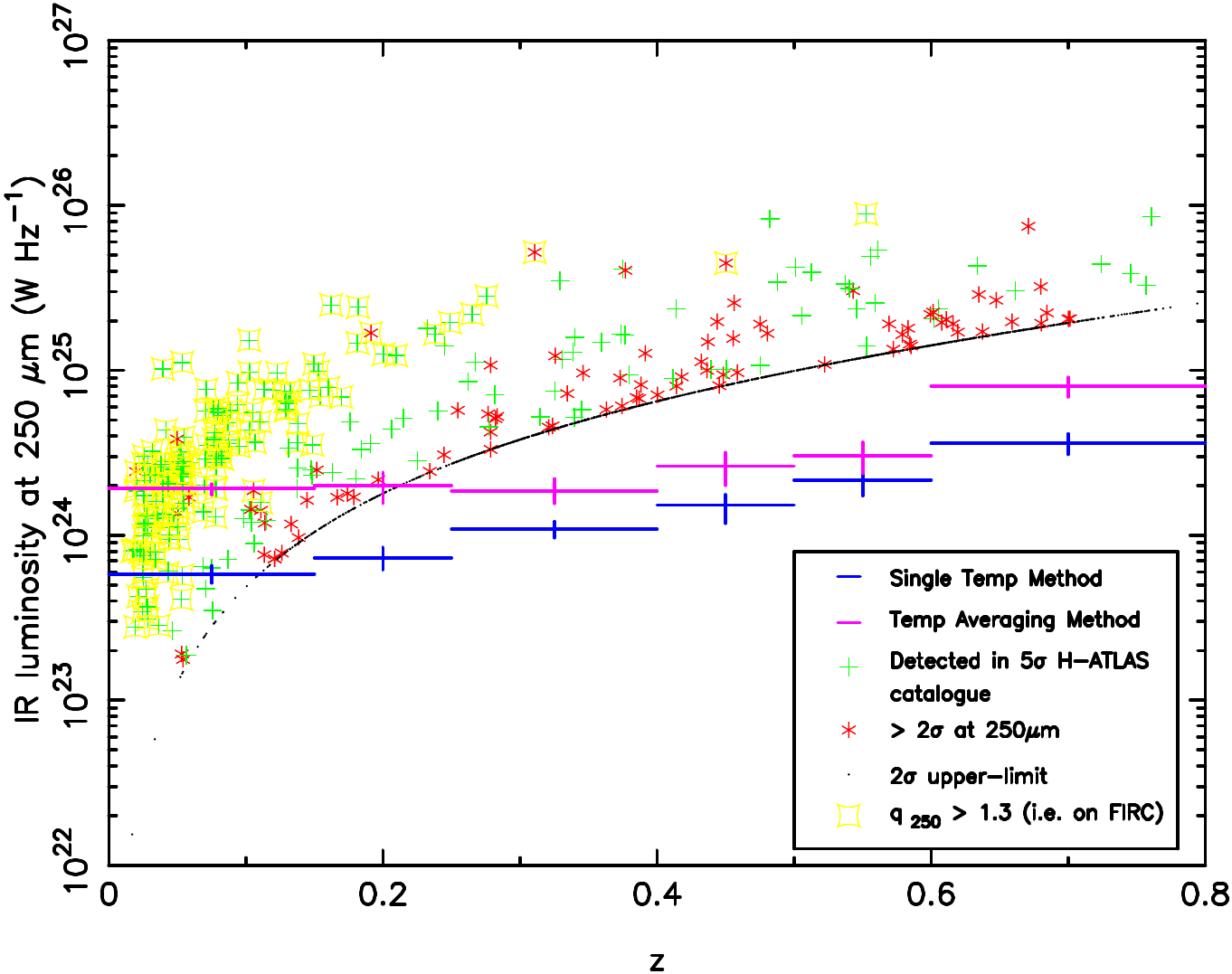}
\includegraphics[scale=0.6,clip,trim=0cm 0cm -0.5cm 0cm]{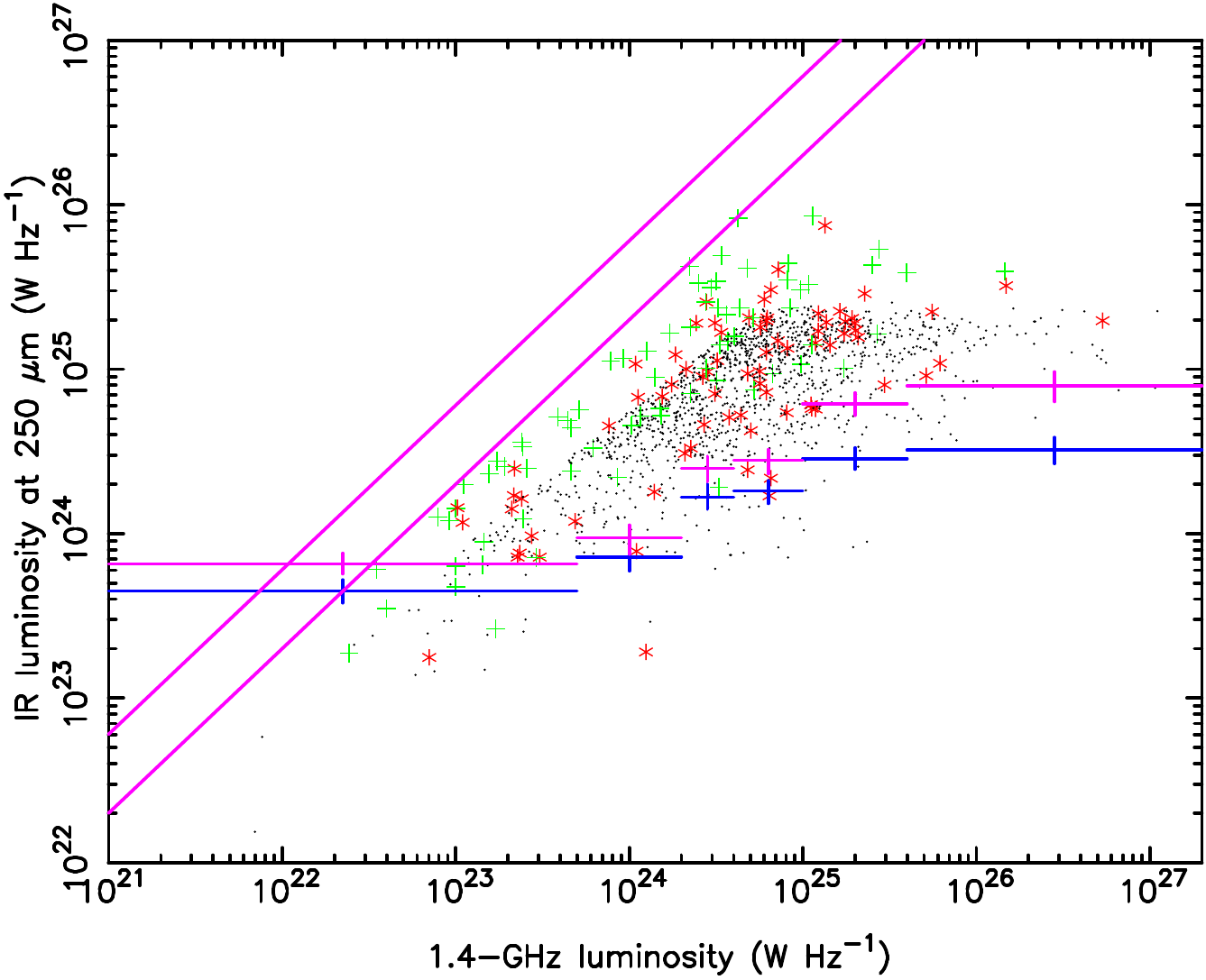}
\includegraphics[scale=0.6]{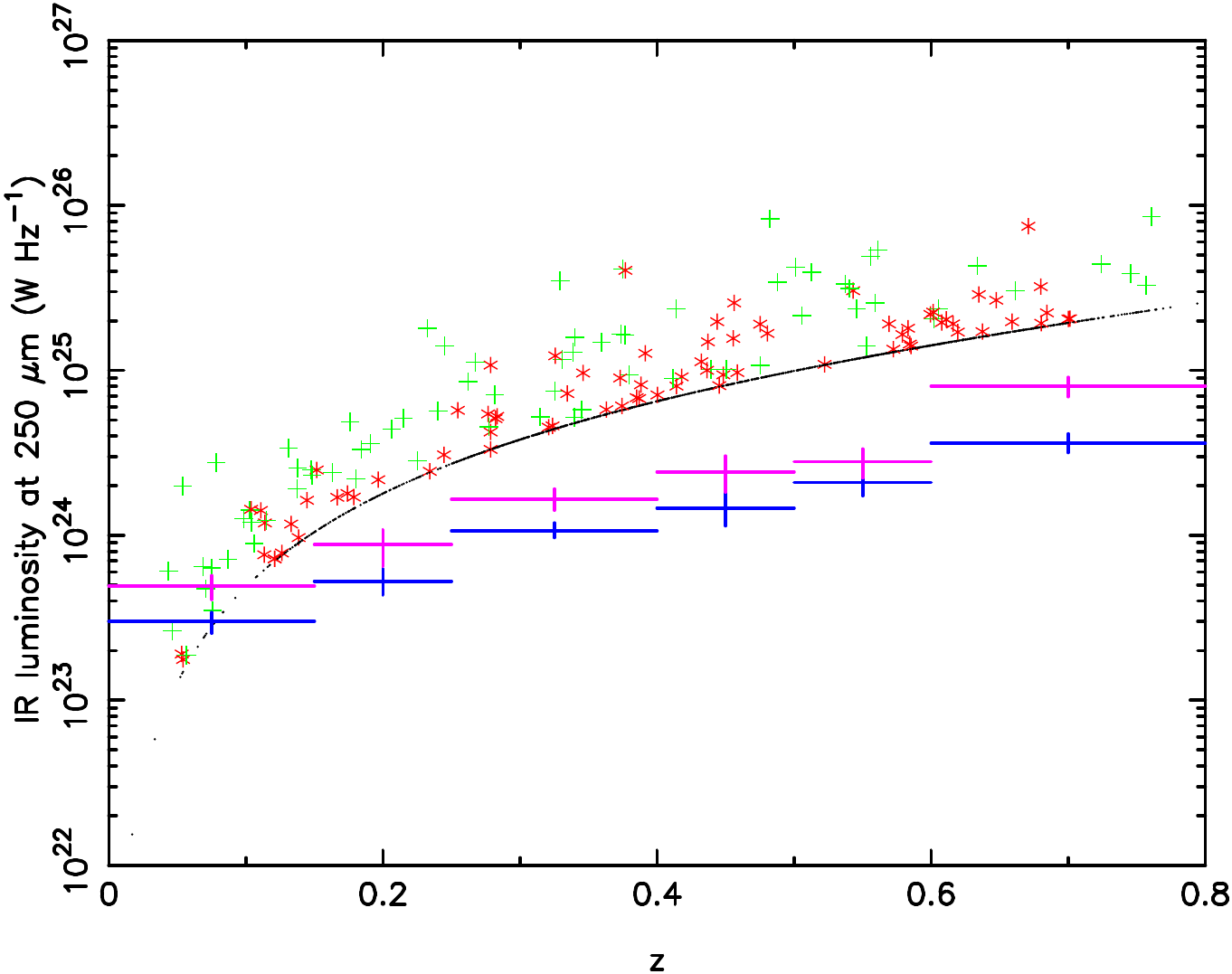}
\caption{Stacked infrared luminosity at 250\,$\mu$m as a function of 
(left) radio luminosity and (right) redshift. (Top) Radio sources on the 
FIRC are included, (bottom) radio sources on the FIRC are not 
included. Plotted sources, which are shown for illustrative purposes 
only, have had their luminosities calculated using the $T=20$\,K isothermal model 
described in the text. 
FIR-luminosities calculated via the single temperature model / temperature averaging
method are in blue / magenta).  Green crosses are 5$\sigma$ detected 
sources defined by the H-ATLAS Phase 1 catalogue whilst red stars and black points 
are the $>$\,2$\sigma$ and 2-$\sigma$ upper-limits, respectively. 
Those sources with $q_{250}>1.3$ are identified with yellow boxes. 
The upper and lower magenta lines are $q_{250}=1.78$ (the FIRC 
as defined by \citealt{Jarvis_2010}) and $q_{250}=1.3$ respectively.}
\label{ir-sf}
\end{figure*}

\subsection{Averaging dust temperatures}
\label{MLT}

In Section\ \ref{l250}, we calculated the luminosity at 250\,$\mu$m using an 
isothermal model with $T=20$\,K. Obviously, such a model contains no 
information on the dust temperature but we know from H13 that a 
wide range of temperatures are observed in the radio selected objects. In order 
to gain some constraints on the temperature, we calculated a weighted average 
of the fitted temperatures of the sources in each bin.
To this end, we take a simple $\chi^2$ approach. We cycle through 
temperatures between 5-55\,K and fit to all the five \emph{Herschel} flux 
bands, allowing only the normalization of each source to vary (assuming 
a modified blackbody SED with $\beta$\,$=$\,1.8). For each temperature 
step, we sum the $\chi^2$ value produced by each source. 
This results in a $\chi^2$ distribution as a function of temperature from 
which we pick the temperature with the lowest $\chi^2$ value.
The errors are calculated by finding the range where $\Delta\chi^{2}=1$.

Although our primary aim is to extract temperature information from each 
bin, we can use this temperature to calculate a K-corrected 
luminosity at 250\,$\mu$m for every source in the bin, again allowing the 
normalization of each source to vary. It is important to stress here that the 
normalization of each source is determined from all five \emph{Herschel} bands, 
so the resulting $S_{250}$ value (and therefore the $L_{250}$) may 
differ substantially from the nominal $S_{250}$ value in the 250\,$\mu$m
map. Confusion bias is likely to affect the 500\,$\mu$m more
strongly than any other due to the increased size of the beam and higher
noise levels. To ensure that our temperatures were not strongly biased by the
inclusion of this band, we computed all the temperature estimates presented
in the paper without the 500\,$\mu$m band and found that the derived 
temperatures were not significantly different.
In the next section we compare the 
$L_{250}$ output of the $T=20$\,K isothermal model to that of the temperature averaging 
method described above.

\subsection{FIR luminosities: Comparison of methods}

In this section we compare the luminosities of the single temperature and 
temperature averaging models. To do this, we calculate the stacked luminosity 
at 250\,$\mu$m of our radio-selected samples as a function of redshift 
and radio luminosity, the results of which are shown in Fig.\ \ref{ir-sf} 
(we note that all the bins are significantly detected with respect to 
the background). The upper plots include the sources on the FIRC 
whilst the lower plots exclude such sources. The luminosities
calculated using the $T=20$\,K isothermal model are shown by the stacks in 
blue while the luminosities calculated via the temperature averaging
method are shown by the stacks in magenta. 
All the plotted sources in Fig.\ \ref{ir-sf} have had their luminosities calculated 
assuming the single temperature model and are shown for illustrative purposes 
only. In the top left panel of Fig.\ \ref{ir-sf}, it is clear that the sources on the 
FIRC are well separated from the radio AGN population. This is not surprising 
given the clean separation observed in Fig.\ \ref{q250}, nevertheless, it is 
encouraging to observe this bi-modality as a function of radio luminosity.

Turning first to the upper plots in Fig.\ \ref{ir-sf}, which contain the 
star-forming sources, we see that the luminosities from both methods 
do not agree particularly well. The temperature averaging method 
produces fitted temperatures ranging from 17-30\,K, with all but the 
highest redshift and radio luminosity stacks having fitted temperatures above 
$T=20$\,K. Furthermore, the luminosities
derived from the temperature averaging method appear systematically 
higher than the ones observed for the single temperature model.
This is because the individual sources in the temperature averaging
method are free to find their own normalizations based on information 
in all the \emph{Herschel} bands. This normalization value is usually 
higher where star-forming objects are included, probably due to the flux
contained in the PACS bands.
If one removes the hot, bright low redshift star-forming objects (Fig.\ \ref{ir-sf}: 
lower plots), the luminosity difference between both methods decreases
significantly. Nevertheless, there are still some systematic differences
between both methods.

When investigating the FIR luminosities of radio-detected and non radio-detected 
galaxies, we will not use the temperature averaging method. This is because  
under that method, the 250\,$\mu$m luminosity of any given source is dependent on 
the luminosities of others in that bin though the temperature estimation. The $T=20$\,K 
isothermal model described above is not susceptible to such correlations, as a single 
temperature and independent flux density values ($S_{250}$) are used when calculating 
individual luminosities.
The systematic differences observed between methods suggest
that the errors on the absolute luminosities of both methods have probably 
been underestimated, since they are highly dependent on the chosen method. 
However, our science conclusions will depend on the \emph{relative} luminosity 
between samples (i.e between radio-detected and non radio-detected galaxies). Therefore, 
in so far as $T=20$\,K is a reasonable temperature estimate for the K-correction 
of our sample, the single temperature method described in Section\ \ref{l250} is sufficient
for the investigation of luminosity differences between samples. All the 
following plots have had their IR luminosities calculated using the isothermal
$T=20$\,K model, while the temperatures have been generated using the 
temperature averaging method.

\subsection{What does the luminosity at 250\,$\mu$m trace?}
\label{trace}

\begin{figure*}
\includegraphics[scale=0.6,clip,trim=0cm 0cm -0.5cm 0cm]{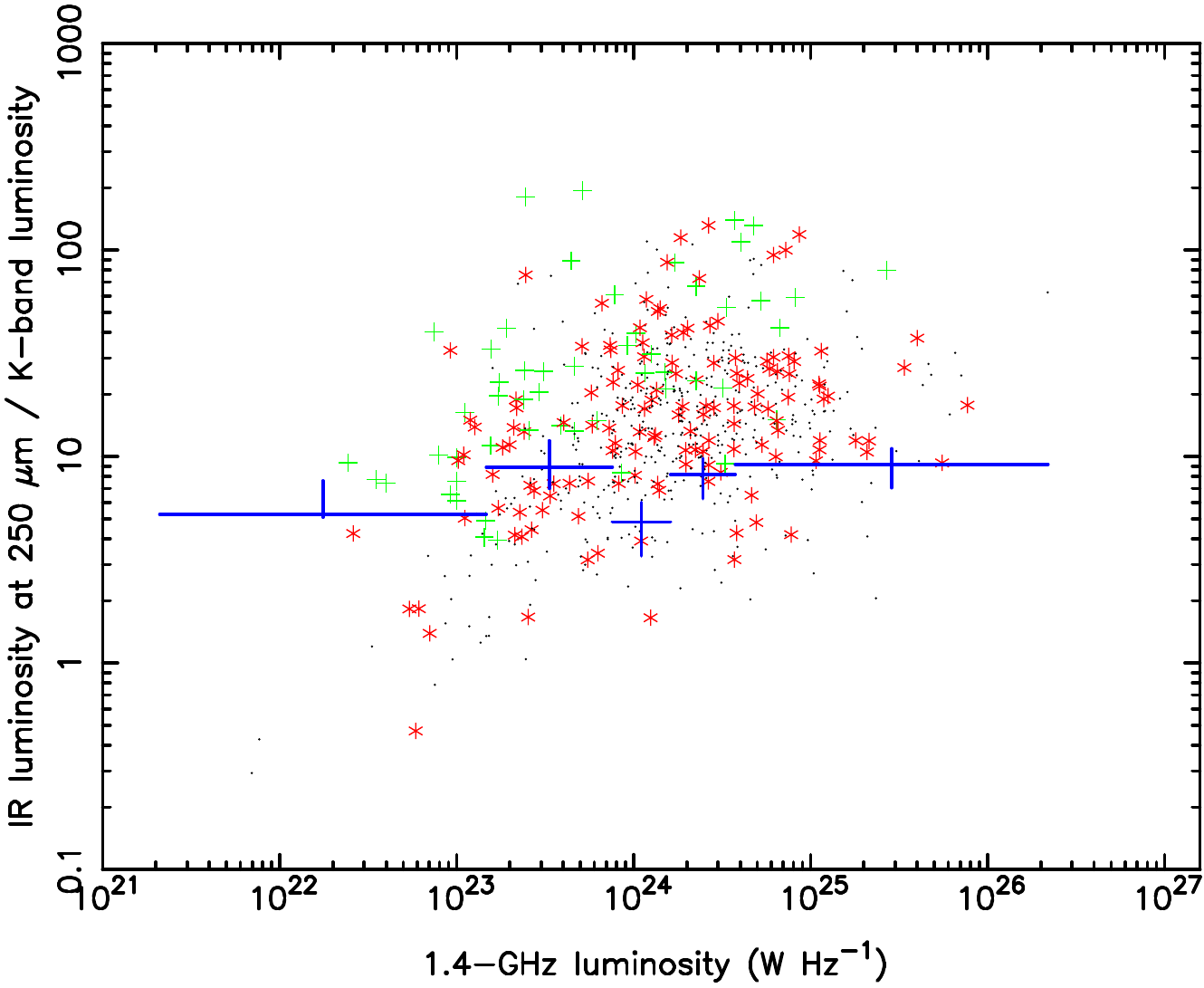}
\includegraphics[scale=0.6,clip,trim=0cm 0cm 0cm 0cm]{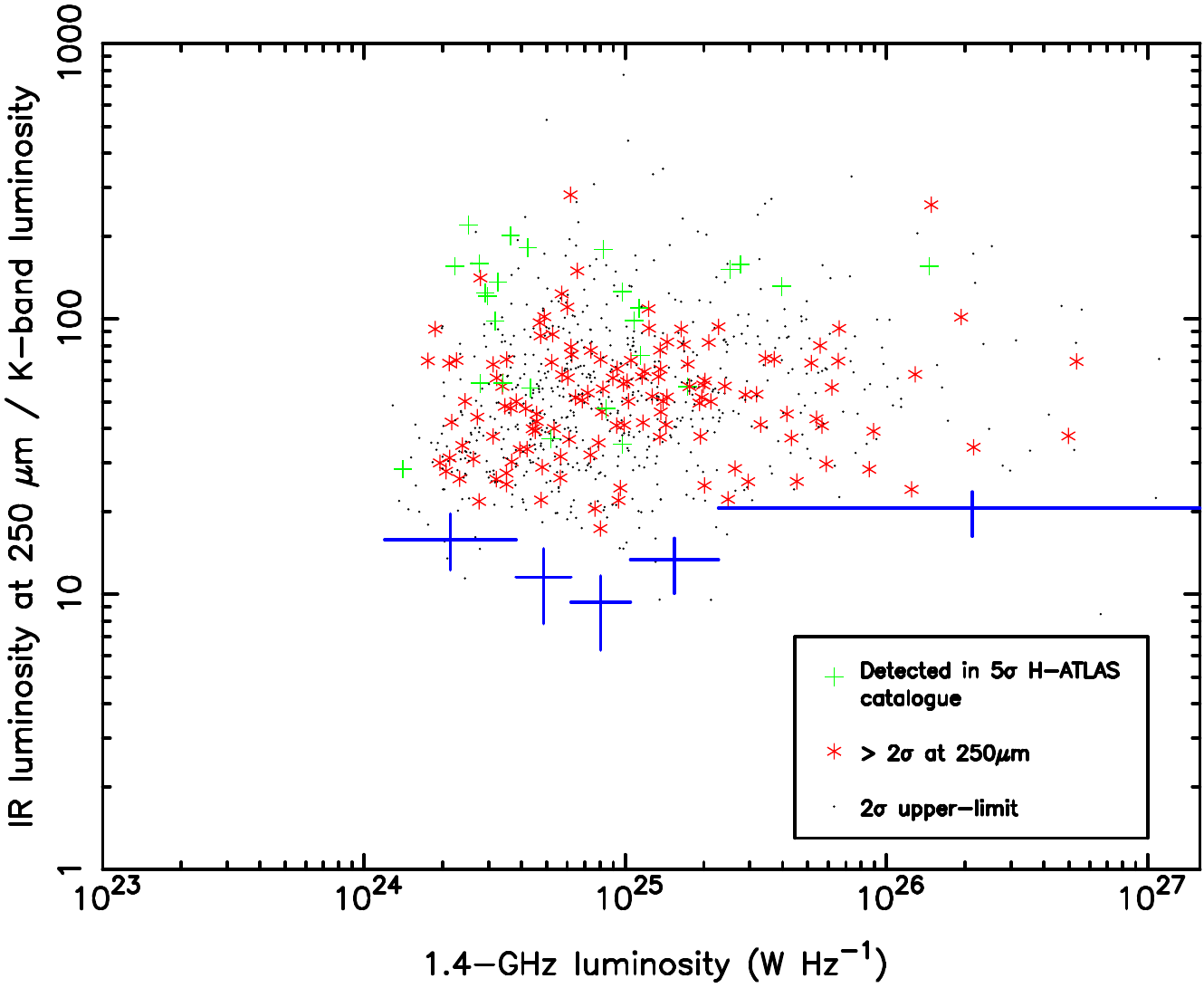}
\caption{Stacked FIR luminosity at 250\,$\mu$m (W\,Hz$^{-1}$) divided by the K-band
luminosity (W\,Hz$^{-1}$), as a function of 1.4-GHz radio luminosity for sources $z<0.4$ (left) 
and $z>0.4$ (right). Sources on the FIRC are not included.
Note the different axial ranges present in both plots.}
\label{fir-rad-nosf}
\end{figure*}

\begin{figure*}
\includegraphics[scale=0.6,clip,trim=0cm 0cm -0.5cm 0cm]{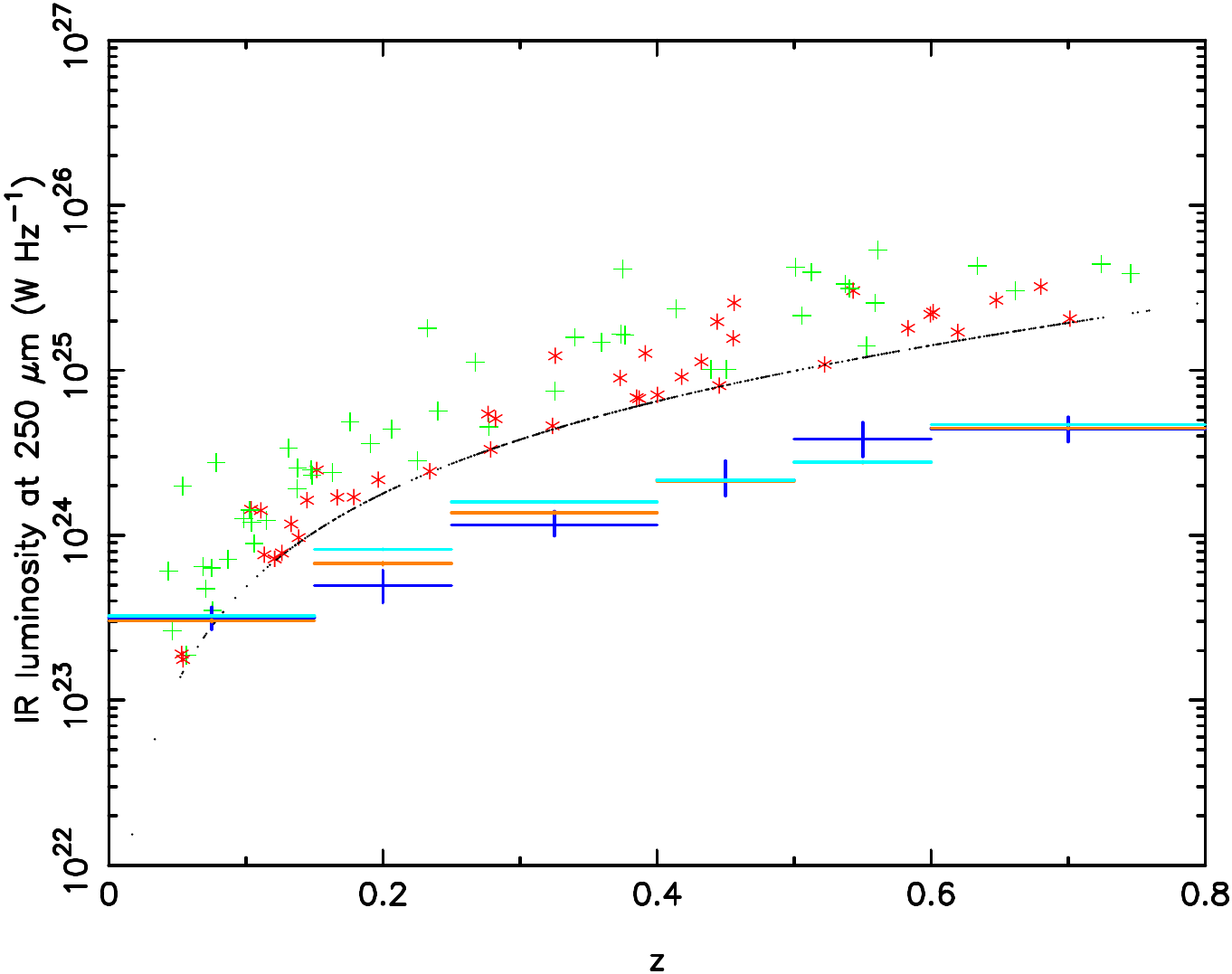}
\includegraphics[scale=0.6,clip,trim=0cm 0cm 0cm 0cm]{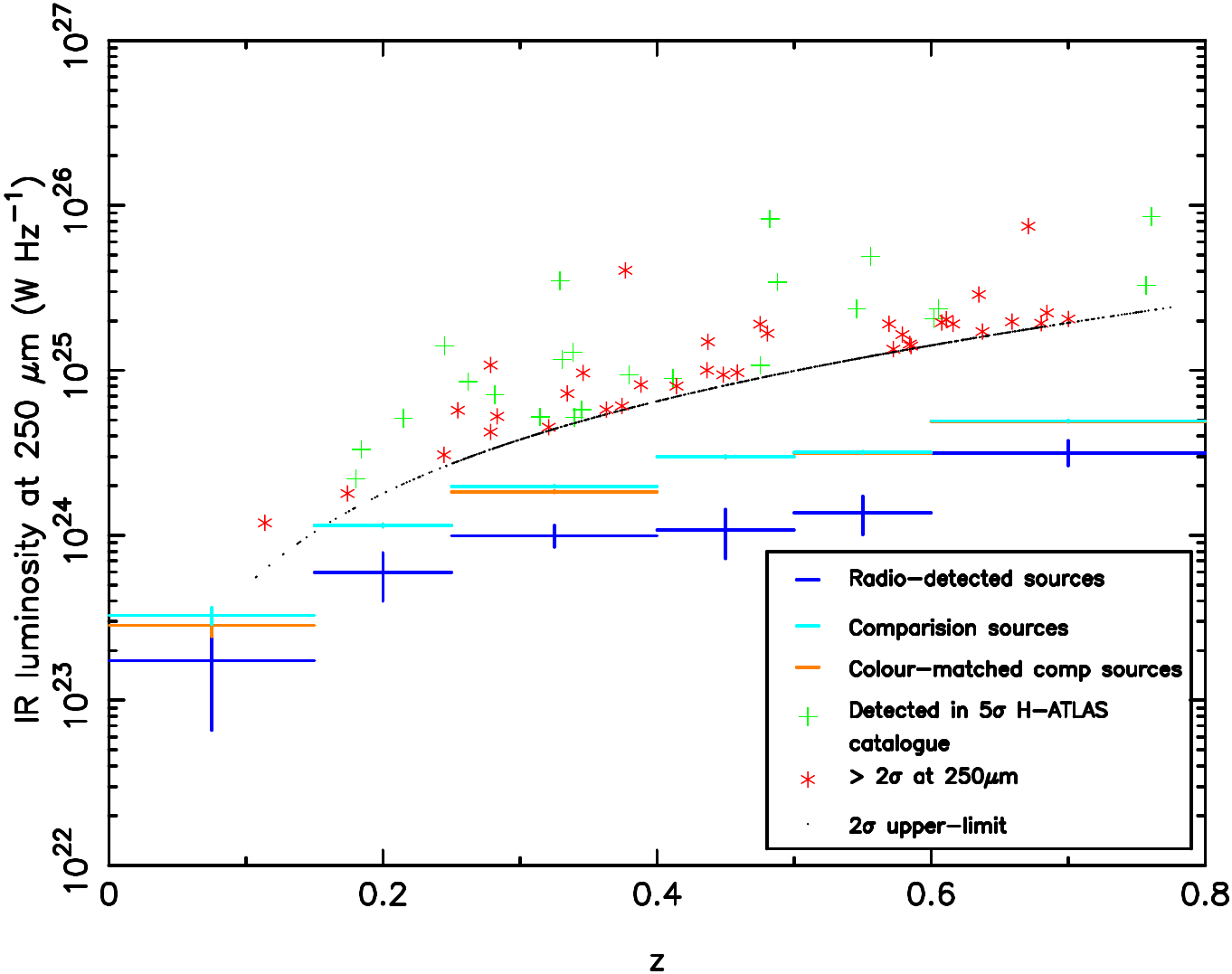}
\caption{Stacked FIR luminosity at 250\,$\mu$m as a function of redshift. 
Stacks shown in blue, cyan and orange are from the radio-detected, 
non radio-detected and g$'$-\,r$'$ colour matched non radio-detected galaxy samples, respectively.
Left and right plots contain the sub-1.5\,$L_{K}^{*}$ and super-1.5\,$L_{K}^{*}$ samples,
respectively. Sources on the FIRC are not included. In some redshift bins, the colour-matched
non radio-detected stacks are not clearly visible because the stacked luminosities are very similar to 
the unmatched non radio-detected luminosities.}
\label{fir-rad-z-nosf}
\end{figure*}

Using the rest-frame luminosity at 250\,$\mu$m minimises
the dispersion in our FIR luminosity calculations, but what does it physically 
trace? First, we must tackle the issue of AGN contamination from 
(1) synchrotron emission and (2) hot dusty tori. The first of these
points has largely been addressed in the text above: point sources which
may be face-on quasars were filtered from the catalogue at an early stage
and sources which have radio spectral indices below 0 are expected to 
make up less than 2 per cent of the radio source catalogue (see 
Section\ \ref{rad-lum}).
As for dusty tori surrounding AGN, they peak in the mid-IR. 
When the required mid-IR data are available, which is not the case for our 
sample at present, decompositions of the SEDs of 
radio-detected and non radio-detected AGN tend to show that observer-frame Herschel SPIRE 
bands are dominated by cool dust rather than by the torus component, even for 
powerful AGN with luminous tori (e.g. \citealt{Barthel_2012}; \citealt{DelMoro_2013}). 
Since the luminosity calculations rely solely on the 250\,$\mu$m band, we 
do not expect any AGN contamination. In our temperature averaging method,
we use PACS 100\,$\mu$m data which at $z=0.8$ (the maximum redshift in our
sample) corresponds to a rest-frame wavelength of 55.5\,$\mu$m. Although there is
some potential for AGN contamination here we note that broadly, the torus luminosity 
scales with the radio luminosity (\citealt{Hardcastle_2009}), and the vast majority of our 
sources are below the threshold for `powerful radio galaxies' ($\sim$1$\times$10$^{26}$\,WHz$^{-1}$)
where bright tori might be expected. In addition, the noise in the PACS bands mean that
PACS photometry is lightly weighted relative to the SPIRE bands in the temperature 
fitting method. Therefore, we do not believe that our luminosities or temperatures have 
been significantly affected by AGN components.

In reality, dust exists at a range of temperatures whose integrated
luminosity is often fitted by one, or two, modified isothermal blackbodies 
(e.g. \citealt{Dunne_2011}). 
This has been driven by an inability to determine, and thus constrain, the 
physical composition and\,/\,or temperature distribution of the dust, which 
would allow more sophisticated models to be used. Nevertheless, for the 
purposes of understanding what the luminosity at 250\,$\mu$m represents 
in our sample, it is instructive to assume a two-temperature blackbody model. 
Such models attempt to separate warm dust, which is expected to dominated 
the IR luminosity output, from the cold dust, which is expected to dominate the 
dust mass (\citealt{Dunne_2011}). 

In general, priors for the cold and warm components range from 10-25\,K 
and 30-60\,K respectively (e.g. \citealt{da_cunha_2008}). At 250\,$\mu$m, 
both components may contribute to the total luminosity. 
For example, assuming a warm dust component with a temperature of 32\,K 
and a cold dust component with a temperature of 15\,K, with the latter contributing 
90 per cent of the total dust mass (typical for sources detected in H-ATLAS; 
\citealt{Dunne_2011}), we find that both components would contribute equally to 
$L_{250}$. Although other temperature combinations are possible within the range 
of temperature priors quoted above, this demonstrates our inability to state 
unambiguously that $L_{250}$ is dominated by either warm or cold dust.
This degeneracy is important, as it is likely that the warm dust 
is the component that traces heating by on-going star-formation. 
Given the quality of our data, we make no attempt to partition 
$L_{250}$ into a warm and cold component, although, in Section\
\ref{FIR-size}, we attempt to use the temperatures generated from 
the temperature averaging method to shed light on this degeneracy.


\section{Results}
\label{results}

\subsection{FIR properties of the K-band luminosity-separated radio galaxies}
\label{FIR-mass}

Galaxy mass is a fundamental indicator of host galaxy properties. 
This is especially true of radio galaxies. There is increasing evidence 
for fundamental mass-dependent differences in the radio luminosity 
functions, accretion rates and possibly fuelling mechanisms of radio 
galaxies (c.f. \citealt{Best_2005b}; \citealt{Best_2012}). For these reasons, 
we investigate the luminosity at 
250\,$\mu$m of our radio AGN as a function of the K-band luminosity 
(a proxy for stellar mass), as well as redshift and radio luminosity. 

In the lower left-hand panel of Fig.\ \ref{ir-sf}, it is clear that $L_{250}$ increases
with the 1.4-GHz radio luminosity. However, such a relation may be the
result of observational selection effects. 
For example, the incompleteness that arises from the flux-limited 
nature of the NVSS means that we have not identified low radio-luminosity
sources at high redshift. The inclusion of such sources would likely flatten the
observed relation since we expect these sources to have systematically
higher $L_{250}$ values than their lower redshift counterparts (the
FIR luminosity of the galaxy population as a whole increases with redshift 
for the redshift range under discussion; \citealt{Dunne_2011}). 
A similar select effect is at work with respect to the K-band magnitude-limited 
LAS images. Such a limit means that we do not detect low-luminosity 
(i.e. low-mass) radio sources at high redshift (see Fig.\ \ref{kz}). 
In order to remove any stellar-mass bias, we divide $L_{250}$ by $L_{K}$ in 
Fig.\ \ref{fir-rad-nosf}. This new value can be thought of as the specific FIR 
luminosity at 250\,$\mu$m. 
To limit the redshift bias, we plot the sources below
and above $z=0.4$ in the left and right-hand panels of Fig.\ \ref{fir-rad-nosf}, 
respectively. The sizes of the stacked bins were chosen so as to ensure that an 
equal number of radio sources were in each stack. Due to the flux limited nature 
of our radio-selected sample, we do not have any sources below 
1$\times$10$^{24}$\,WHz$^{-1}$ in the $z>0.4$ sample. 
The best fit (assuming a line with a gradient of zero) specific FIR luminosity 
at 250\,$\mu$m of the $z<0.4$ and $z>0.4$ samples are 6.5$\pm$1.7 and
13.4$\pm$3.9 respectively.
Nevertheless, it is clear that any relation between $L_{250}$ and 1.4-GHz radio 
luminosity disappears after these corrections, although the FIR luminosities of the
$z>0.4$ sources are systematically higher than that of the $z<0.4$ sources. This implies 
that the strength of a galaxy's radio emission has little or no bearing on its specific FIR 
luminosity at 250\,$\mu$m.

\begin{figure}
\begin{center}
\includegraphics[scale=0.22,clip,trim=1.5cm 1cm 0cm 0cm]{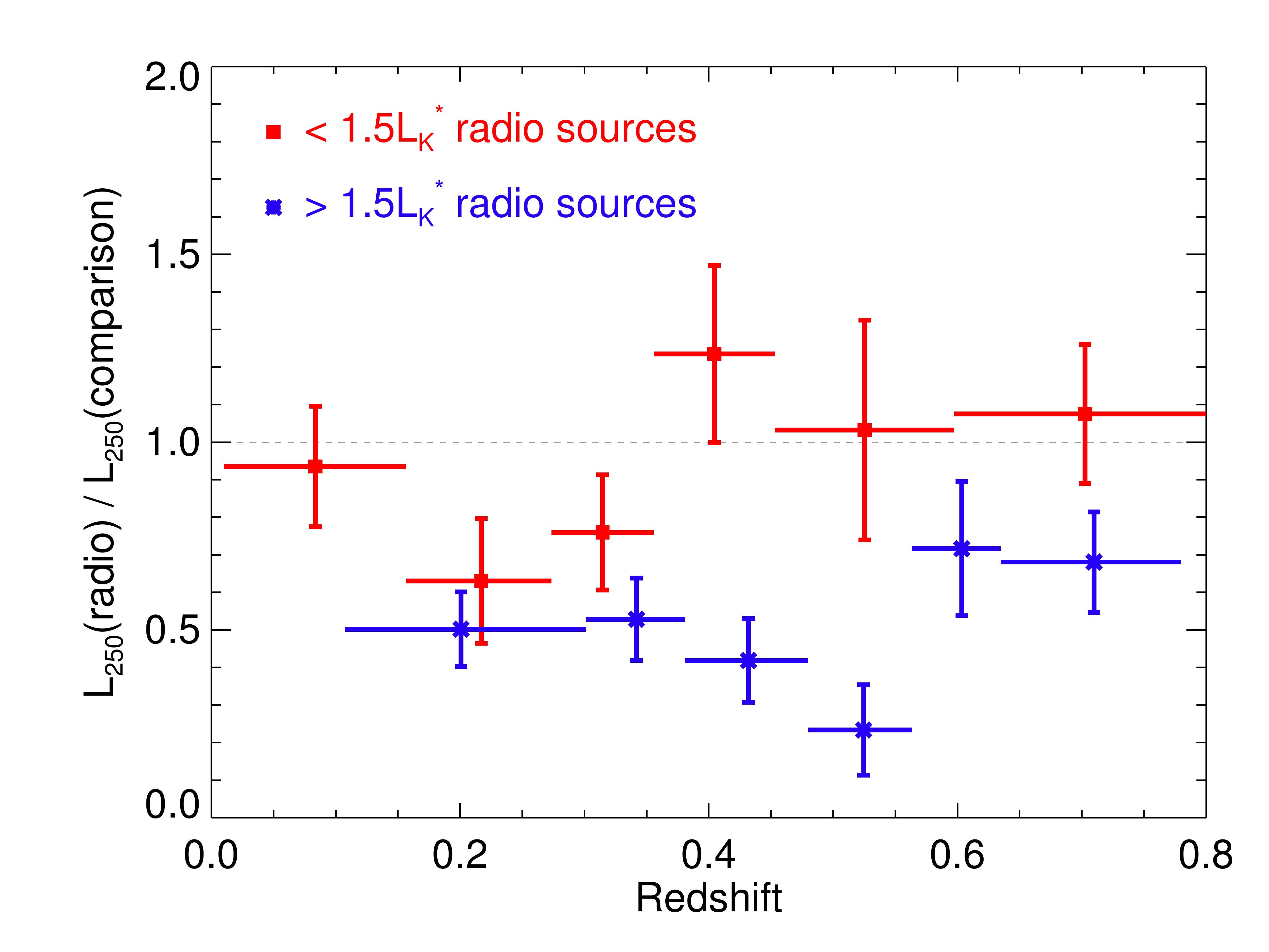}
\caption{Plotted is the ratio of the stacked $L_{250}$ values
of radio-detected and non radio-detected galaxies against redshift.
The radio sample is divided into sub (red) and super (blue) 
1.5\,$L_{K}^{*}$ samples. Stack sizes are chosen to ensure each 
bin contains the same number of radio sources.}
\label{l250-fraction}
\end{center}
\end{figure}

Next, we compare the IR luminosities of the radio-detected sources against
their non radio-detected counterparts. Here, it is important to ensure that any FIR 
luminosity deviation between the samples is not simply the result of 
differing optical colours.
The average K-corrected g$'$\,-\,r$'$ colours of the sub-1.5\,$L_{K}^{*}$ and 
super-1.5\,$L_{K}^{*}$ radio-detected samples are 0.83 and 0.90 respectively,
while average colours of the sub-1.5\,$L_{K}^{*}$ and 
super-1.5\,$L_{K}^{*}$ non radio-detected samples are 0.75 and 0.91 respectively. 
Thus, while the average colours of the super-1.5\,$L_{K}^{*}$ samples
match up well, the sub-1.5\,$L_{K}^{*}$ radio-detected galaxies have
redder colours than their non radio-detected counterparts.
 In order to take colour into account, we filter our sub and super-1.5\,$L_{K}^{*}$ 
comparison samples so as to match the distribution of g$'$\,-\,r$'$ colours observed
in the sub and super-1.5\,$L_{K}^{*}$ radio-detected samples. 
For each subsample, we do this by first imposing a global g$'$\,-\,r$'$ colour cut, defined 
by the middle 98 per cent of the radio galaxy colour distribution; this removes any outliers 
in the radio-detected and comparison samples. Next, we randomly discard one per cent of 
the galaxies in the comparison sample and then run a K-S test to establish how well the
new colour distribution matches to the radio-detected distribution. Discarding a \emph{different} 
one per cent 
and repeating the process $N$ times, where $N$ is the number of comparison sources
in the sample, allows us to select the best-matching \emph{reduced} catalogue. After selecting 
the best matching catalogue, we again discard one per cent of the remaining sources 
in the comparison sample and repeat the entire process. This is done until the null 
hypothesis probability returned by the K-S test exceeds 10 per cent. This colour-matching process 
is performed for the sources in each redshift bin separately. The main disadvantage of such 
a method is the potential loss of large fractions of the original comparison sample. We 
lose $\sim$\,55 per cent of the comparison sources in the sub-1.5\,$L_{K}^{*}$ sample;
however, this still leaves us with over 136,000 sources. In the super-1.5\,$L_{K}^{*}$
comparison sample, we only lose $\sim$\,6 per cent of the sources 
leaving us with over 26,000 sources. 
In Fig.\ \ref{fir-rad-z-nosf}, we plot $L_{250}$ versus
redshift. The radio galaxy stacks are plotted in dark blue whilst the comparison 
galaxy stacks are plotted in cyan. Sources on the FIRC have been excluded from
this plot. The stacks shown in orange are generated using the colour matched 
comparison samples. Hereafter, `colour-matched' samples refer to the comparison 
samples which have undergone the matching process described above.

It is immediately apparent from Fig.\ \ref{fir-rad-z-nosf} that both the stacked radio 
and comparison samples show an increase in $L_{250}$ with redshift. This is 
consistent with the picture that dust masses and star-formation activity were higher 
in the past (e.g. \citealt{Dunne_2011}). 
Turning to the differences between the K-band luminosity separated radio-galaxies, 
there is a suggestion that the sources in the sub-1.5\,$L_{K}^{*}$ sample have higher FIR 
luminosities than those in the super-1.5\,$L_{K}^{*}$ sample. In order to quantify this difference,
we divide the $L_{250}$ values of the sub-1.5\,$L_{K}^{*}$ sample, shown in 
Fig.\ \ref{fir-rad-z-nosf}, by the equivalent $L_{250}$ values of the super-1.5\,$L_{K}^{*}$
sample.
We find the sub-1.5\,$L_{K}^{*}$ radio-detected sample to be 1.53$\pm$0.32 
times as luminous as the super-1.5\,$L_{K}^{*}$ radio-detected sample. Therefore, 
although there is some evidence for higher FIR luminosities in the 
sub-1.5\,$L_{K}^{*}$ sample, this result is not significant due to the large errors.

\begin{figure*}
\includegraphics[scale=0.6,clip,trim=0cm 0cm -0.5cm 0cm]{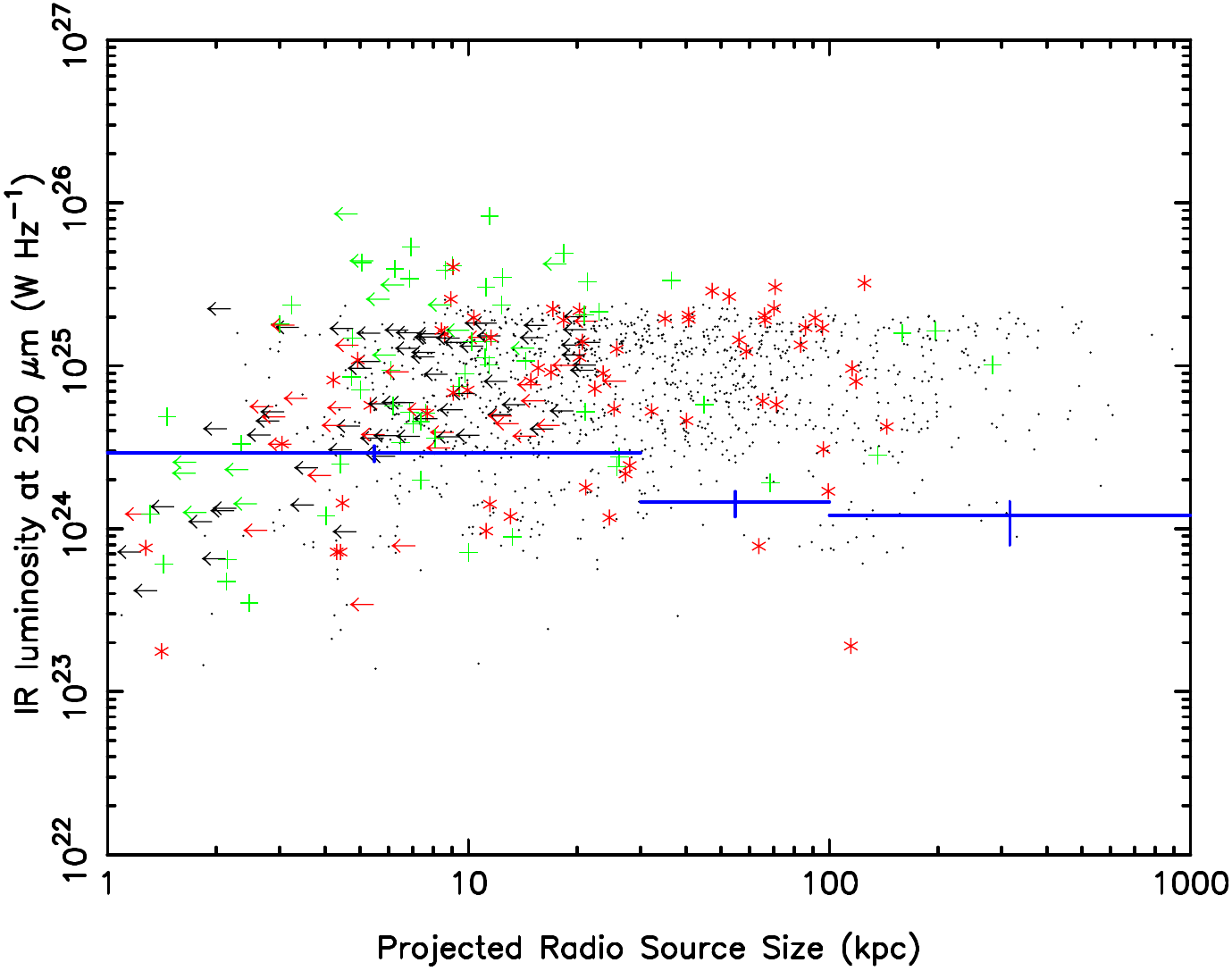}
\includegraphics[scale=0.6,clip,trim=0cm 0cm 0cm 0cm]{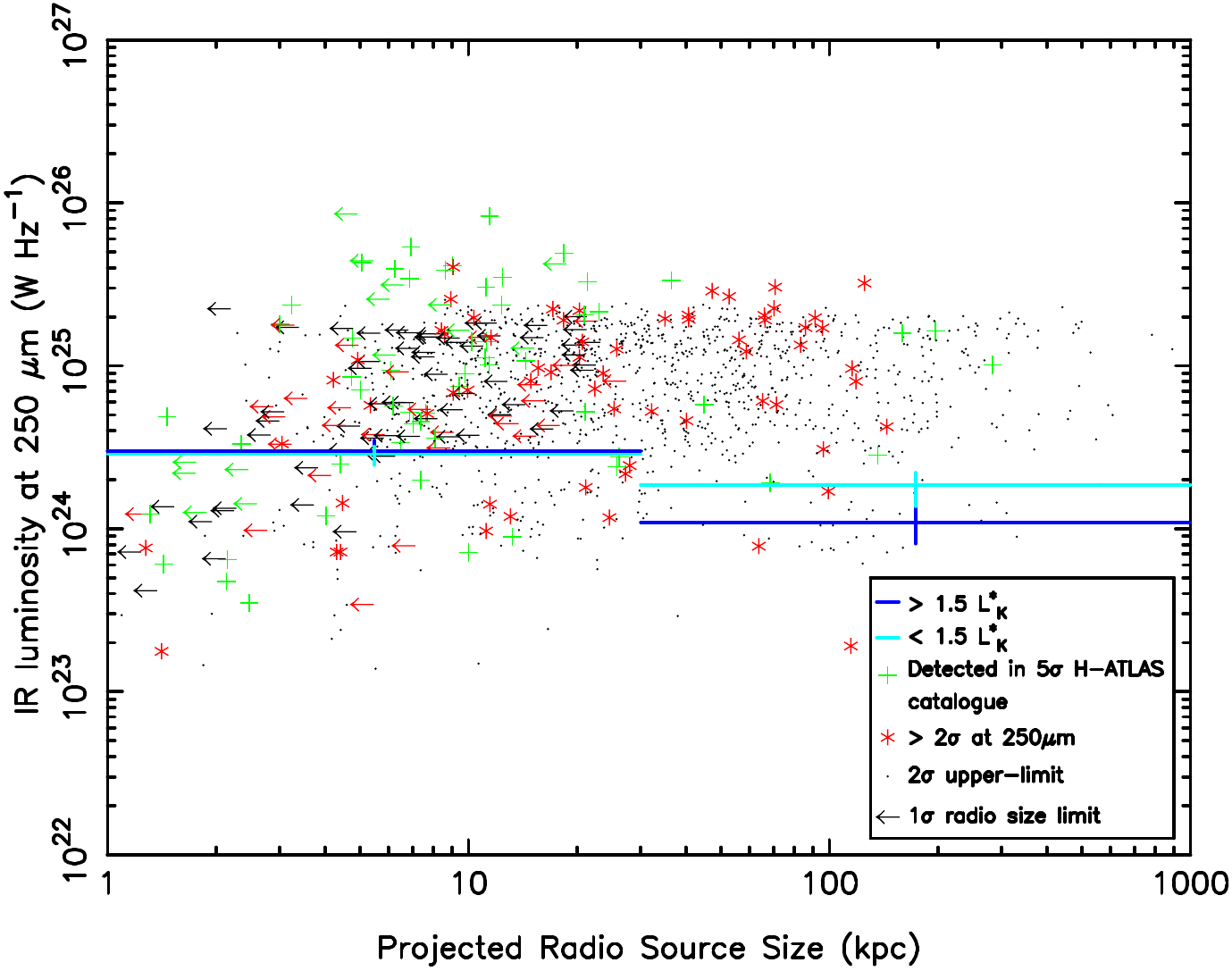}
\caption{FIR luminosity at 250\,$\mu$m versus the projected 
radio source size. Radio sources on the FIRC are excluded. 
On the left-hand panel, we plot the stacked values for the entire 
radio-detected sample. On the right-hand panel, we plot the stacked 
values for the sub (cyan) and super (blue) 1.5\,$L_{K}^{*}$ 
radio-detected samples. Plotted sources are shown for illustrative 
purposes only. Green crosses are 5$\sigma$ detected sources 
defined by the H-ATLAS Phase 1 catalogue whilst red stars / black 
points are the $>$\,2$\sigma$ / 2-$\sigma$ upper-limits respectively. 
The arrows represent the upper limits on the sizes of the 
radio sources.}
\label{lrs}
\end{figure*}

In order to judge, quantitatively, whether there is any difference between the 
radio-detected and comparison samples, we divide the $L_{250}$ values of 
the the radio-selected stacks by the colour-matched comparison stacks and 
show the results in Fig.\ \ref{l250-fraction}. The sizes of the stacked bins were 
chosen so as to ensure that an equal number of radio sources were in each 
stack. However, it is important to note several significant caveats before interpreting 
the results. 
Firstly, because the NVSS is a flux-limited survey, we do not have high-redshift, 
low-luminosity radio sources in our sample: i.e, as we move to higher redshifts,
we are preferentially selecting powerful radio AGN and placing any 
low-luminosity radio galaxies in the comparison sample. Secondly, since our 
base galaxy catalogue is magnitude limited in the $K$ and $r'$ bands, we become
progressively incomplete with increasing redshift, losing faint and\,/\,or low-mass 
sources. Nonetheless, it is important to note that the radio and comparison 
samples have been selected from the same parent catalogue. 

Fitting a line with a 
gradient of zero, we find the best fitting line for the sub-1.5\,$L_{K}^{*}$ 
radio-detected sample is 0.89$\pm$0.18 times the FIR luminosity at 
250\,$\mu$m of the colour matched comparison sample. Thus, being 
radio-detected is not associated with higher luminosities at 250\,$\mu$m. 
This is markedly different from what we see with the super-1.5\,$L_{K}^{*}$ 
sample, which is significantly better constrained. 
We find the best fitting line for this sample to be  
0.49$\pm$0.12 times the FIR luminosity at 250\,$\mu$m of the colour 
matched comparison sample. Thus, it is clear that super-1.5\,$L_{K}^{*}$ 
radio-galaxies show a strong 250\,$\mu$m luminosity deficit with respect 
to their non radio-detected counterparts. We discuss the significance of the above 
results in Section\ \ref{discussion}.

\subsection{FIR emission and radio source sizes}
\label{FIR-size}

In Fig.\ \ref{lrs} we plot $L_{250}$ versus the projected radio source size. The 
size of the radio source can be thought of as a \emph{rough} proxy for the age 
of the radio source (e.g. \citealt{Kaiser_1997}). Sources on the FIRC have been 
excluded from this plot. The sizes of the stacked bins were chosen so as to keep 
all the radio sources with poorly constrained projected sizes in 
the first bin (i.e. $<$\,30\,kpc). All the bins are significantly detected on a K-S test with respect to 
the background. On the left-hand panel of Fig.\ \ref{lrs}, we plot the results using 
the entire radio-detected sample with radio sizes. There is a clear indication that the 
luminosity at 250\,$\mu$m 
falls with increasing source size. The null hypothesis, that $L_{250}$ is independent 
of radio size, is rejected at the 95 per 
cent confidence level. However, we must first discount the possibility that this fall in FIR
luminosity is due to observational selection effects. Such an effect could be 
orchestrated if `compact' radio sources ($<$\,30\,kpc) were at higher redshifts
than their more `extended' counterparts ($>$\,30\,kpc), during which FIR luminosities 
of all sources were higher, or at higher radio 
luminosities where the FIR luminosities \emph{of this sample} are higher (see left-hand 
side of Fig.\ \ref{ir-sf}). However, the sub-30\,kpc radio sources are 
slightly biased towards the low-redshift and low radio-luminosity ends compared to the
super-30\,kpc sample. 

To help elucidate the origin of the FIR excess at small radio source sizes, we calculate
the `best-fit' temperature, obtained using the temperature averaging method described in 
Section\ \ref{MLT}, for compact and extended radio sources. 
We find the fitted temperatures of compact and extended radio galaxies 
are 27.1$\pm$0.3\,K and 13.1$\pm$0.3\,K respectively.
The higher temperatures are probably due to enhanced inter-stellar radiation 
fields (ISRF) heating cool dust. The source of this heating is likely to be
enhanced levels of, possibly jet-induced or merger associated star-formation.

On the right-hand panel of Fig.\ \ref{lrs}, we separate the sample using the
$K$-band luminosity. The super-1.5\,$L_{K}^{*}$ and sub-1.5\,$L_{K}^{*}$ radio 
galaxy stacks are shown in blue and cyan, respectively. Due to the 
sparsity of extended radio sources in the sub-1.5\,$L_{K}^{*}$ sample, we only
plot two stacks divided at 30\,kpc. It is clear that both samples show a fall
in $L_{250}$ with increasing radio source size, although the super-1.5\,$L_{K}^{*}$ 
sample dominates this trend. We discuss the implications of these results
in Section\ \ref{discussion}.

\begin{figure}
\includegraphics[scale=0.6,clip,trim=0cm 0cm 0cm 0cm]{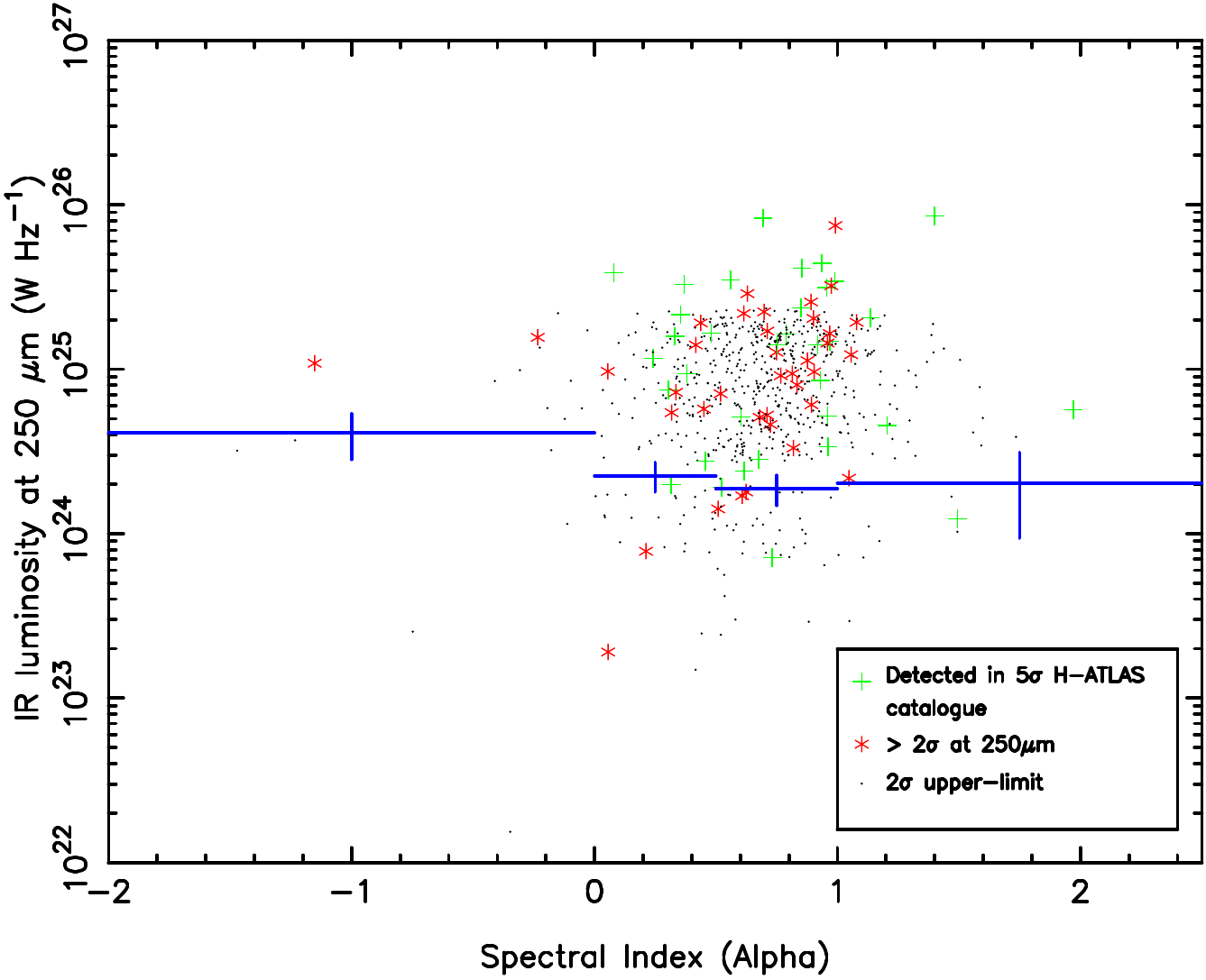}
\caption{FIR luminosity at 250\,$\mu$m versus the radio spectral index 
($S \propto \nu^{-\alpha}$) for the subset of NVSS-detected sources 
(1.4-GHz) with GMRT detections (330\,MHz).}
\label{lalpha}
\end{figure}

\subsection{FIR emission and spectral indices}

The large survey at 330\,MHz performed by the GMRT over the H-ATLAS 
Phase 1 fields (Mauch et al. in preparation) allows us to determine whether 
spectral indices are related to FIR luminosity. Spectral indices can be thought
of as another \emph{rough} proxy for radio source age.
This is because, during the lifetime of a radio source, the electrons with the 
highest energies in a homogeneous magnetic field will radiate energy at a 
faster rate than their lower energy counterparts. This leads to a steepening of
the spectral index at high frequencies. Thus, we might expect a negative relation 
between FIR luminosity and spectral index, if the radio source is somehow 
synchronised with the onset of star-forming activity. 

In Fig.\ \ref{lalpha}, we plot $L_{250}$ versus spectral index for the subset of radio 
sources with detections at 330\,MHz. Excluding flat spectrum sources whose FIR 
emission may be contaminated by non-thermal emission ($\alpha<0$), 
we find no evidence for any relation between 
spectral index and $L_{250}$ \emph{in our sample} of radio galaxies. The lack of a 
negative relation implies that either the star-forming and radio AGN activity are
not well (or at all) synchronised in the current sample of radio galaxies or that
spectral indices do not represent a good proxy for the age of our radio galaxies.
It is also possible that the frequency range over which the spectral index is calculated 
is too low to see a relation between electron ageing and star formation. 
The addition of further data at radio frequencies, with which the shape of the spectral 
distribution could be reconstructed and thus modelled, may allow one to distinguish 
between the cases described above.


\section{Discussion}
\label{discussion}

The most obvious trend observed in both the stacked radio-detected and 
non radio-detected galaxy samples is the increase in the rest-frame 
luminosity at 250\,$\mu$m with redshift. This is at least qualitatively 
consistent with previous results suggesting that dust masses and star-formation 
rates evolve strongly with redshift (e.g. \citealt{Dunne_2011}).

Perhaps the most interesting result from Section\ \ref{results} is the different 
behaviour of the sub and super-1.5\,$L_{K}^{*}$ radio-detected galaxies with respect 
to their K-band and g$'$-\,r$'$ matched non radio-detected galaxies. 
While the 250\,$\mu$m luminosity of the sub-1.5\,$L_{K}^{*}$ radio-detected sample
is indistinguishable from its matched comparison sample, the super-1.5\,$L_{K}^{*}$
radio-detected sample shows a strong FIR luminosity deficit with respect to its matched 
comparison sample. However, what does a deficit at 250\,$\mu$m tell 
us about the physical properties of the host systems?

The answer may depend on the fitted isothermal dust temperature. If the dust 
temperature of two matched samples are similar, any differences in the FIR 
luminosity will likely be due to differing dust masses. However, if the dust temperature 
of both samples differ greatly, then any differences in the FIR luminosity will likely 
be due to differing SFRs.
To calculate the dust temperature of the radio-detected and comparison samples,
we use the temperatures generated by the temperature averaging method described in 
Section\ \ref{MLT}. The primary drawback of using this method is that a small number
of bright, and possibly warm sources may dominate the $\chi^{2}$ fitting, resulting in higher 
temperatures than one would expect by simple averaging. This implies that a minority of 
sources may be responsible for the majority of the temperature difference observed between two
samples.\\

\begin{figure*}
\begin{center}
\includegraphics[scale=0.22,clip,trim=0.0cm 1cm 0cm 0cm]{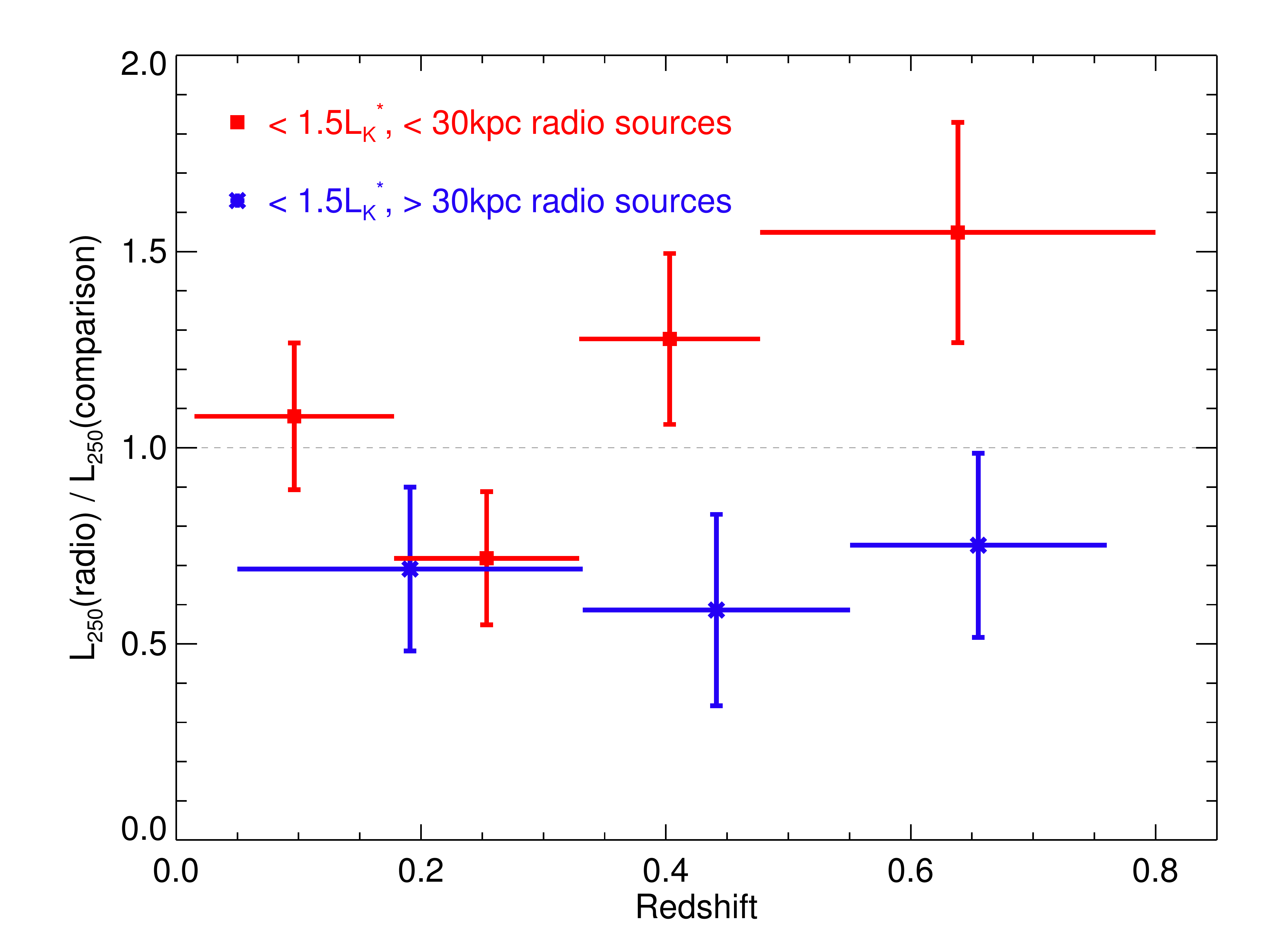}
\includegraphics[scale=0.22,clip,trim=0cm 1cm 0cm 0cm]{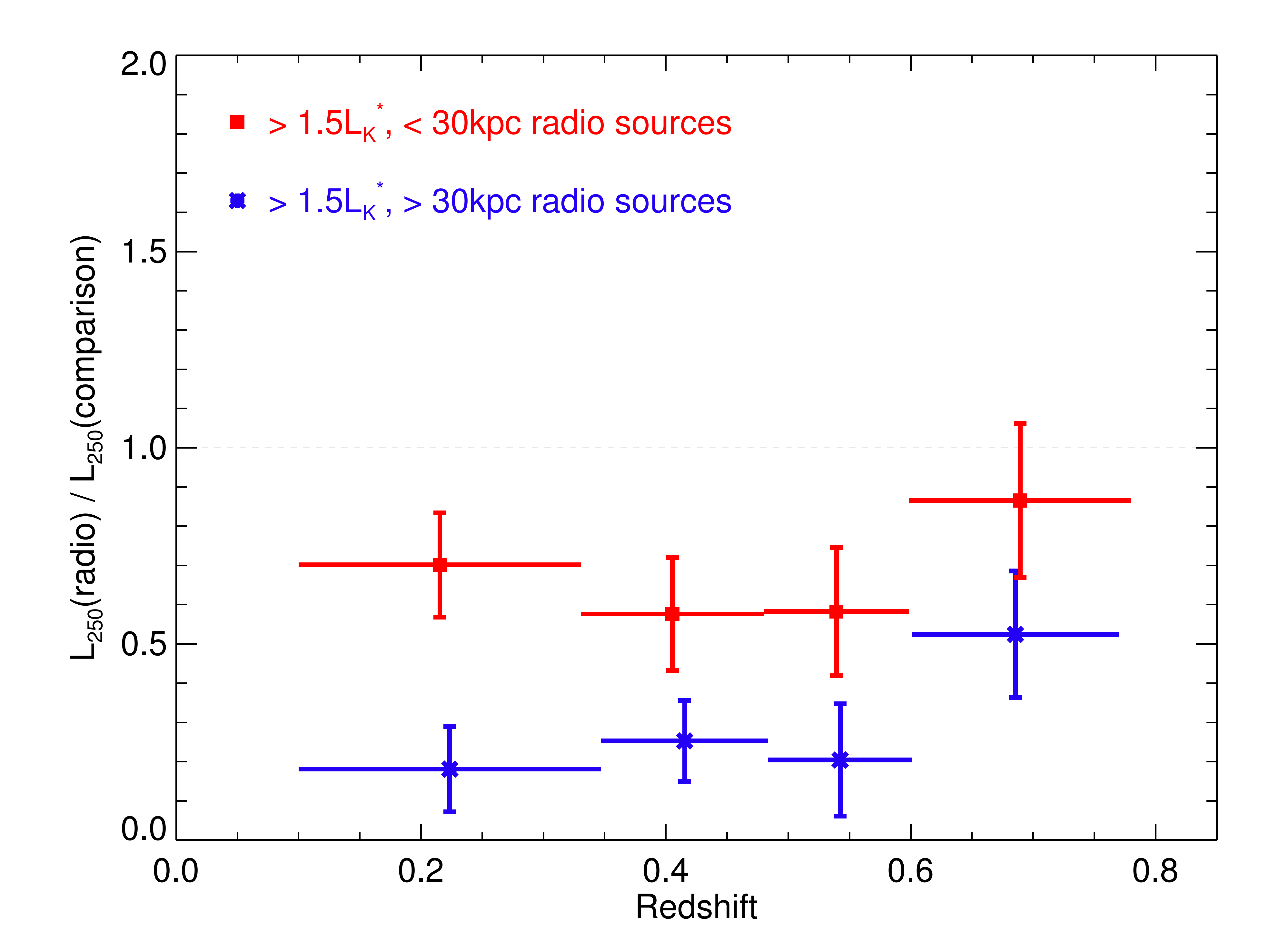}
\caption{Stacked luminosities at 250\,$\mu$m of the radio-detected samples divided by the 
colour-matched comparison samples. The sub-1.5\,$L_{K}^{*}$ (left) and super-1.5\,$L_{K}^{*}$ 
(right) radio-detected samples are separated into compact ($<$30\,kpc) and extended ($>$30\,kpc)
radio galaxies. The red stacks represent the $<$30\,kpc samples while the blue stacks 
represent the $>$30\,kpc samples.}
\label{l250-fraction-size}
\end{center}
\end{figure*}

Turning first to the sub-1.5\,$L_{K}^{*}$ radio-detected results, we find that the fitted temperatures 
of the radio-detected and non radio-detected samples are 20.5$\pm$0.5\,K and 19.0$\pm$0.2\,K, 
respectively. The combination of similar FIR luminosities and dust temperatures can be
interpreted in several ways. For example it may simply mean that the radio jets, or the 
mechanism that leads to radio emission, are not strongly coupled to the dust reservoirs 
in all sub-1.5\,$L_{K}^{*}$ galaxies. Alternatively, it is possible the radio-detected sample 
contains at least two distinct populations, with different FIR luminosities, which are not
well separated by stellar mass alone. 
As discussed in Section\ \ref{intro}, HERGs, which tend to be hosted in 
lower-mass galaxies, are expected to be triggered as a result of on-going 
gas-rich major mergers. We might then expect the radio sources in this 
sample to show a strong 250\,$\mu$m luminosity excess due to a combination 
of increased star-formation and higher dust masses. However, HERGs dominate 
the radio luminosity function only above $\sim$10$^{26}$ WHz$^{-1}$ 
(\citealt{Best_2012}), well above the luminosity of most of the sources in this sample. 
Thus, the sub-1.5\,$L_{K}^{*}$ radio-detected sample is likely to contain a mixture of 
HERGs and LERGs. This makes it difficult to be certain that the null result is due to 
similar dust masses, as the presence of a small number of HERGs would render this 
interpretation invalid. Nevertheless, if one takes the sub-1.5\,$L_{K}^{*}$ radio-detected 
population as a whole, it is clear that there is no evidence for differing FIR luminosities 
or (at least isothermal) dust temperatures.

The super-1.5\,$L_{K}^{*}$ radio-detected sample, which we expect to be 
dominated by LERGs, shows a clear FIR luminosity deficit when compared 
to its colour-matched comparison sample. The suggestion here is that 
being radio-detected results in \emph{lower} FIR luminosities. The fitted 
temperatures of the radio-detected and comparison sample are 16.2$\pm$0.5\,K 
and 21.2$\pm$0.2\,K respectively, implying that the observed differences 
may be due to a combination of lower dust masses and SFRs. 
One of the most obvious interpretations is that radio-jets directly destroy dust or inhibit
star formation in these galaxies. If this were the case, we might expect a 
negative relation between the mechanical heating of the ISM by the jet (which itself is 
dependent on the 1.4-GHz radio luminosity) and $L_{250}$. However, we do not have
the data to calculate mechanical heating caused by the jet. The lack of a 
negative relation between the specific 250\,$\mu$m luminosity and 1.4-GHz radio 
luminosity, suggests that a causal connection between the radio jet and the 
luminosity deficit at 250\,$\mu$m is unlikely, although evolution in $L_{1.4}$
may have washed out the effect. It is difficult to imagine hot, ionised radio-jets 
coupling efficiently to cold, neutral dust. If such a mechanism was real and 
ubiquitous, we might have expected to also see a $L_{250}$ deficit in the sub-1.5\,$L_{K}^{*}$ 
radio-detected sample. That we do not suggest we must look elsewhere 
for a potential solution.

All massive galaxies sit at the bottom of large, dense and hot potential wells. 
However, what distinguishes radio-detected systems from their non radio-detected 
counterparts is that they inhabit the richest, hottest and densest of these environments. This is 
both because the radio activity is thought to be fuelled by the hot gas from the IGM (see 
Section 1) and because hot gas is required to confine the radio lobes. This hot gas may 
ultimately lead to a suppression of the total dust content 
of any galaxy embedded in this environment. Radio jets are known to drive shocks 
in the halos of their host galaxies, heating them (e.g. \citealt{Croston_2009}). This 
heating has been used to solve the so called `cooling flow problem', where only small 
amounts of cold gas are observed 
to condense out of hot-gas halos despite radiative cooling timescales which 
are much less than the age of the host (\citealt{McNamara_2012}).
This leads to lower star-formation rates and also lower dust production
rates. In addition, the hot gas may directly destroy, or simply suppress the 
formation of dust particles inside the galaxy though sputtering (\citealt{Draine_1979}). 
Further, any in-falling galaxy is likely to be ram-pressure stripped of its gas and dust 
content by the hot gas, limiting the alternative sources of cold gas and dust available to 
galaxies embedded in this environment (see \citealt{Abramson_2011} and 
\citealt{Vollmer_2012} for examples of ram stripping in the Virgo cluster). 
Thus, the tendency of massive ($>$1.5\,$L_{K}^{*}$) radio galaxies to inhabit rich 
environments may explain the implied lower dust masses and SFRs observed in 
these systems and the lack of such a result in the sub-1.5\,$L_{K}^{*}$ 
radio galaxies.\\

Next, we turn to the results based on the radio source size. First, we 
note that the dust temperatures of compact radio sources, whether they 
are in the sub or super-1.5\,$L_{K}^{*}$ sample, are always higher
than in their more extended counterparts. These higher temperatures
are most likely to be the result of increased star-formation activity
in the compact radio sources, a result reminiscent of the findings of
\citet{Dicken_2012}, who showed that various indicators of
star-formation activity were more likely to be seen in compact sources
drawn from their small sample of powerful objects. There are two
possible explanations for such a result. On the one hand, it may be that the small
sources ($<30$ kpc) are energetically coupled to the cold gas in their
host galaxies, and drive star formation directly (i.e. the often
invoked but poorly understood process of `jet-induced star
formation'; see \citealt{Rees_1989}, \citealt{Begelman_1989}, 
\citealt{Gaibler_2012}). 
On the other hand, it may be that these young radio
sources were triggered by an event, such as a merger, which also
triggered a burst of star formation that will have died away by the
time the sources are older and larger. Both models are somewhat at
odds with the picture of LERGs (which are expected to make up the
majority of sources in the super-1.5\,$L_{K}^{*}$ sample) as being 
quiescent systems with little cold gas available for star formation, but it is
possible that the result is driven by a minority of compact HERGs.
In any case, these results indicate an important relationship between 
radio-jet activity and star formation, although whether this relation is 
causal or a result of common correlation with environmental properties is unclear.\\

It is interesting to ask if the environmental explanation for the lower SFRs and dust 
masses of the super-1.5\,$L_{K}^{*}$ radio-detected sample can be confirmed 
using the results based on radio source sizes. This is a difficult question 
to answer as the relevant environmental parameters are gas temperature 
and density, not radio source size (although extended radio sources may 
simultaneously select more extended, hotter and possibly richer environments). 
Nevertheless, we attempt to combine both results in order to see whether
a self-consistent picture can emerge. \\
\indent We combine the analysis by splitting the sub and super-1.5\,$L_{K}^{*}$ samples 
into two bins at 30\,kpc (i.e, as a function of radio source size). 
On the left and right-hand sides of Fig.\ \ref{l250-fraction-size}, we plot the results 
for the sub and super-1.5\,$L_{K}^{*}$ samples relative to their colour-matched 
comparison samples. Stacks shown in red and blue represent the compact and 
extended radio samples, respectively. 
Fitting a line with a gradient of zero to the sub-1.5\,$L_{K}^{*}$ subsamples, we
find that compact and extended radio sources are 1.05$\pm$0.19 and 0.68$\pm$0.25 
times as bright at 250\,$\mu$m as the colour matched comparison sample, with 
fitted temperatures of 22.2$\pm$0.5\,K and 13.5$\pm$0.5\,K respectively. Therefore,
while the temperature difference is significant with high confidence, the luminosity 
difference is not, due to large errors.
Repeating the process with the super-1.5\,$L_{K}^{*}$ subsamples, we
find that compact and extended radio sources are 0.66$\pm$0.15 and 0.26$\pm$0.12 
times as bright at 250\,$\mu$m as the colour-matched comparison sample, with 
fitted temperatures of 22.0$\pm$1.0\,K and 12.2$\pm$0.5\,K, respectively.\\
\indent From these results, it is clear that the $L_{250}$ deficiency in super-1.5\,$L_{K}^{*}$ 
radio-detected galaxies is driven primarily by extended sources, with low dust 
temperatures. 
This is not altogether surprising, since it was obvious from the right-hand side
of Fig. \ref{lrs} that extended super-1.5\,$L_{K}^{*}$ radio-detected galaxies had
lower FIR luminosities than their compact counterparts. Such a result is consistent
with an environmental explanation for the suppression of $L_{250}$, as larger 
radio sources should, on average, select larger and denser halos than smaller radio sources.
These results also suggest that if one accounts for the, presumably temporary jet-induced
star-formation of compact radio sources, \emph{all} massive radio sources have intrinsically 
lower FIR luminosities and temperatures, and therefore dust masses, than their non 
radio-detected counterparts. Our results suggest that massive radio-detected galaxies 
should have about 25 per cent of the FIR luminosity of comparable non radio-detected 
galaxies. 
This result is inconsistent with a direct causal link between the jet and the 
suppression of $L_{250}$, as one might expect lower 250\,$\mu$m luminosities in 
compact sources relative to extended sources.

In order to verify that the environment regulates the FIR properties of radio 
galaxies, a careful study of the environment and its relation to the host's dust content
and star-formation activity is needed. These results suggest 
that the FIR luminosities and temperatures of radio galaxies depend on the 
stellar mass as well as the radio source size. Understanding the origin of these 
dependencies will be useful for galaxy evolution models which often use the 
radio activity of galaxies to quench galaxy growth.


\section{Conclusions}
\label{conclusions}

Using a large sample of radio galaxies selected in the \emph{Herschel}-ATLAS 
Phase 1 area, we investigate the FIR properties of the radio AGN population
between $0.01<z<0.8$ by stacking the 100, 160, 250, 350 and 500\,$\mu$m 
H-ATLAS maps in order to obtain stacked rest-frame luminosities 
at 250\,$\mu$m and dust temperatures. The main results are:
\begin{enumerate}

\item We find no relation between the rest-frame luminosity at 250\,$\mu$m 
divided by the K-band luminosity (i.e. the specific $L_{250}$), and the 1.4-GHz 
radio luminosity in our sample of radio-detected galaxies. These results imply 
that a galaxy's \emph{nominal} radio luminosity has little or no bearing on the 
star-formation rate and\,/\,or dust mass content of the host system. We stress that
this does not imply that being radio-loud has no indirect or indeed direct
effect on the FIR luminosity of the host system, just that the 
specific $L_{250}$ has no dependence on the nominal value of the radio luminosity.
A more suitable variable such as the work done on the ISM by the jet may yet show 
some sort of relation.

\item The rest-frame luminosity at 250\,$\mu$m
of the radio-detected and non radio-detected galaxies rises with increasing
redshift. This is consistent with the view that most galaxies show
signs of increased dust masses and star-formation activity at higher
redshifts.

\item Compact radio sources ($<$\,30\,kpc) are associated with higher 250\,$\mu$m 
luminosities than their more extended ($>$\,30\,kpc) counterparts. The fitted temperatures
of compact and extended radio galaxies are 27.1$\pm$0.3\,K and 13.1$\pm$0.3\,K 
respectively, implying that an increased dust temperature in compact objects is responsible 
for this observation. This temperature difference suggests that there may be enhanced 
levels of star-formation in compact objects. However, whether this star-formation is it directly 
(e.g. jet induced) or indirectly associated (e.g. merger driven) with the AGN activity
is not yet clear.

\item For matched samples in $L_{K}$ and g$'$-\,r$'$, sub-1.5\,$L_{K}^{*}$ 
radio-detected galaxies have 0.89$\pm$0.18 times the 250\,$\mu$m luminosity
of the comparison sample, with fitted temperatures of 20.5$\pm$0.5\,K and 
19.0$\pm$0.2\,K, respectively. Thus, taken as a whole, there is no evidence
that sub-1.5\,$L_{K}^{*}$ radio-detected galaxies have different FIR properties 
to that of their non radio-detected counterparts.

Splitting the radio sample at 30\,kpc, we find that
the compact and extended radio-detected galaxies have 1.05$\pm$0.19 and 
0.68$\pm$0.25 times the 250\,$\mu$m luminosity of the sub-1.5\,$L_{K}^{*}$ 
non radio-detected comparison sample, and have fitted temperatures of 22.2$\pm$0.5\,K 
and 13.5$\pm$0.5\,K respectively. Thus, although there is a suggestion of a FIR luminosity 
dependence on radio source size, large errors make it difficult to make this statement 
quantitatively. However, the observed temperature difference does suggest that compact 
radio sources may have slightly higher SFRs than their more extended counterparts.

\item For matched samples in $L_{K}$ and g$'$-\,r$'$, super-1.5\,$L_{K}^{*}$ 
radio-detected galaxies have 0.49$\pm$0.12 times the 250\,$\mu$m luminosity
of the comparison sample, with fitted temperatures of 16.2$\pm$0.5\,K and 
21.2$\pm$0.2\,K, respectively. These results imply that  super-1.5\,$L_{K}^{*}$ 
radio-detected galaxies have lower dust masses and SFRs than their non radio-detected 
counterparts.

Splitting the radio sample at 30\,kpc, we find that
the compact and extended radio-detected galaxies have 0.66$\pm$0.15 and 
0.26$\pm$0.12 times the 250\,$\mu$m luminosity of the super-1.5\,$L_{K}^{*}$ 
comparison sample, and have fitted temperatures of 22.0$\pm$1.0\,K and 
12.2$\pm$0.5\,K, respectively.

\item No relation between spectral index and the luminosity at 250\,$\mu$m is
found for the subset of 1.4-GHz radio sources with detections at 330\,MHz. 
The lack of a negative relation implies that 
either star formation and radio AGN activity are not well (or at all) 
synchronised in the current sample of radio galaxies, or that spectral 
indices do not represent a good proxy for the age of our radio sources.
Selecting a sample of merger-driven radio galaxies in which the FIR luminosity 
is dominated by on-going star formation (e.g. HERGs) may demonstrate whether 
any such relation exists in potentially `synchronized' radio sources.

\end{enumerate}

We explain the primary result, the $L_{250}$ deficit seen only in the $>1.5$\,$L^*_K$ radio 
galaxies relative to their colour corrected comparison sample, as an environmental phenomenon. 
Here, the suppression of $L_{250}$ is due to the selection of LERG-like sources which are likely
embedded in the hottest, densest and richest haloes, which the presence of radio jets selects for. 
The hot-gas environment acts to suppress the creation rate of dust particles in the ISM and\,/\,or 
actively destroy dust through sputtering. Ram-pressure stripping of any gas and dust rich galaxy 
falling into the halo would also limit external sources of gas and dust for these systems.
This approach has the 
virtue of explaining why a $L_{250}$ deficit is seen in the $>1.5$\,$L^*_K$ radio galaxies and not 
in the $<1.5$\,$L^*_K$ radio galaxies, and why the $L_{250}$ deficit might be more pronounced 
in extended rather than compact radio sources (i.e. we are selecting larger halo environments). 
In addition, it is consistent with the lack of a relation between the specific $L_{250}$ and the 
1.4-GHz radio luminosity, as the radio jets will tend to keep the halo environment hot, 
with the long timescales for energy transfer washing out any direct relation. It also avoids timescale
issues, which plague any attempt to directly link radio jets with positive or negative feedback. One 
result which is difficult to account for is the systematically hotter temperatures in compact radio sources. 
This suggests enhanced SFRs in compact vs extended radio sources. However, under an 
environmental explanation, there is no reason to suppose a small amount of SF may be stimulated 
by the radio jets for short periods of time. Thus, the environment of radio galaxies is presented as a 
possible explanation for these results as it is the most consistent explanation.
If one assumes that the higher FIR luminosities of compact
radio galaxies are transient, as the FIR luminosities of extended radio galaxies
suggest they are, then the FIR luminosity of \emph{all} massive radio galaxies must 
be intrinsically lower than that of their non radio-detected counterparts, implying 
systematically lower dust masses and SFRs. Thus, massive radio-detected galaxies
may have only about 25 per cent of the FIR luminosity of their non radio-detected 
counterparts.\\

In this work, we have clearly demonstrated that the FIR properties of 
radio galaxies is not identical to that of the non radio-detected population. There
is a strong FIR luminosity dependence on the stellar mass (when compared to
non radio-detected equivalents) and a strong FIR luminosity and temperature 
dependence on the radio-source size.
These conclusions are in disagreement with our earlier results from
H10, where the FIR luminosities of radio-detected and non radio-detected galaxies
were found to be indistinguishable. However, the small sample used in 
H10 (our sample is more than eight times larger) prevented an analysis 
of the FIR properties of radio-galaxies as a function of stellar mass or 
radio source size. In addition, we have added temperature averaged 
estimates and PACS data in order to better interpret the origin of the 
FIR luminosity at 250\,$\mu$m. 
We have attempted to explain our observations in terms of the special
halo environment likely to be present around massive radio galaxies. 
If the environment plays a determinant role, we suggest 
that this is only possible because the continuous heating provided by multiple 
outbursts of radio activity. This may ultimately create the conditions for 
systematically lower dust masses in radio galaxies. In order to prove this, a 
detailed study of the halo environment of radio galaxies and how it relates to 
the FIR properties of 
the host system, is required. The answer to these questions will be of 
great importance for models which use the radio activity of galaxies to
quench galaxy growth.

The work presented here is only able to investigate sources between redshifts 
of $0.01<z<0.8$ because of the constraints imposed via the optical and NIR limits 
from the SDSS and LAS. In the near future, VISTA VIKING data, which is $\sim$1.4 
mag. deeper than the LAS in the $K$-band, will become available, allowing more 
radio galaxies to be classified, increasing the sample size and extending the 
redshift range. Spectroscopic redshift campaigns, which target many of the sources 
not detected in the NIR, will help characterise the FIR\,/\,sub-mm properties of powerful 
high-$z$ radio galaxies. Spectroscopic information can also be useful for radio source
classifications, and in a companion paper (\citealt{Hardcastle_2013}), we have
determined the FIR properties of a well defined sample of HERGs and LERGs.
The work presented here uses less than 25 per cent of the full H-ATLAS survey area, 
implying that large gains in statistics can still be made.

\section*{Acknowledgements}

We thank the anonymous referee for comments which have helped
improve the paper.
{\it Herschel} is an ESA space observatory with science instruments
provided by European-led Principal Investigator consortia and with
important participation from NASA. U.S. participants in {\it
 Herschel}-ATLAS acknowledge support provided by NASA through a
contract issued from JPL. GAMA is a joint European-Australasian project 
based around a spectroscopic campaign using the Anglo-Australian 
Telescope. The GAMA input catalogue is based on data taken from the 
Sloan Digital Sky Survey and the UKIRT Infrared Deep Sky Survey. 
Complementary imaging of the GAMA regions is being obtained by a 
number of independent survey programs including GALEX MIS, VST KIDS, 
VISTA VIKING, WISE, \emph{Herschel}-ATLAS, GMRT and ASKAP providing 
UV to radio coverage. GAMA is funded by the STFC (UK), the ARC 
(Australia), the AAO, and the participating institutions. The GAMA website 
is http://www.gama-survey.org/ . JSV thanks the STFC and RAL for a 
studentship. MJJ acknowledges support from an RCUK fellowship.\\

\bibliographystyle{mn2e}
\bibliography{references.bib}

\end{document}